\documentclass[pdflatex,sn-mathphys-num]{sn-jnl}

\usepackage{graphicx}%
\usepackage{multirow}%
\usepackage{amsmath,amssymb,amsfonts}%
\usepackage{amsthm}%
\usepackage{mathrsfs}%
\usepackage[title]{appendix}%
\usepackage{xcolor}%
\usepackage{textcomp}%
\usepackage{manyfoot}%
\usepackage{booktabs}%
\usepackage{algorithm}%
\usepackage{algorithmicx}%
\usepackage{algpseudocode}%
\usepackage{listings}%
\usepackage{ulem}

\theoremstyle{thmstyleone}%
\theoremstyle{thmstyletwo}%

\theoremstyle{thmstylethree}%

\begin{document}

\title[Article Title]{Thermodynamic sampling of materials using neutral-atom quantum computers}


\author*[1]{\fnm{Bruno} \sur{Camino}}\email{b.camino@ucl.ac.uk}

\author*[2]{\fnm{Mao} \sur{Lin}}\email{maolinml@amazon.com}

\author[3]{\fnm{John} \sur{Buckeridge}}\email{j.buckeridge@lsbu.ac.uk}

\author[1]{\fnm{Scott M.} \sur{Woodley}}\email{scott.woodley@ucl.ac.uk}

\affil*[1]{\orgdiv{Chemistry Department}, \orgname{University College London}, \orgaddress{\street{20 Gordon St}, \city{London}, \postcode{WC1H 0AJ}, \country{United Kingdom}}}

\affil[2]{\orgname{Amazon Braket}, \orgaddress{\city{Seattle}, \postcode{98170}, \state{WA}, \country{United States}}}

\affil[3]{\orgdiv{School of Engineering}, \orgname{London South Bank University}, \orgaddress{\street{103 Borough Rd}, \city{London}, \postcode{SE10 AA}, \state{United Kingdom}}}

\abstract{\unboldmath Neutral-atom quantum hardware has emerged as a promising platform for programmable many-body physics. In this work, we develop and validate a practical framework for extracting thermodynamic properties of materials using such hardware. As a test case, we consider nitrogen-doped graphene. Starting from Density Functional Theory (DFT) formation energies, we map the material energetics onto a Rydberg-atom Hamiltonian suitable for quantum annealing by fitting an on-site term and distance-dependent pair interactions.
The Hamiltonian derived from DFT cannot be implemented directly on current QuEra devices, as the largest energy scale accessible on the hardware is two orders of magnitude smaller than the target two-body interaction in the material. {\color{black}To overcome this limitation, we introduce a rescaling strategy based on a single parameter, $\alpha_v$, which ensures that the distribution sampled by the hardware is well described by Boltzmann-like weights corresponding to those of the material at an effective temperature $T’ = \alpha_v T$, where $T$ is the device sampling temperature.} This rescaling also establishes a direct correspondence between the global laser detuning $\Delta_g$ and the grand-canonical chemical potential $\Delta\mu$.
We validate the method on a 28-site graphene nanoflake using exhaustive enumeration, and on a larger 78-site system where Monte Carlo sampling confirms preferential sampling of low-energy configurations.
}

\keywords{Thermodynamic sampling, Neutral-atom quantum computing, Quantum annealing, Rydberg Hamiltonian, Boltzmann sampling}



\maketitle

\section*{Introduction}
The majority of technologically relevant materials exhibit some degree of disorder, such as vacancies, interstitial species, amorphous regions, solid solutions, or substitutional alloys. 
To predict their properties reliably, it is necessary to employ not only accurate energetic models but also representations capable of capturing the effects of configurational complexity.
Here, we are motivated by the general problem of modelling solid solutions, \textit{i.e.}, predicting the atomic order or disorder for a known lattice where multiple atomic species can occupy equivalent lattice sites.
The result will depend on the temperature and the relative chemical potential between the competing species.

Employing a unit cell with periodic boundary conditions significantly reduces the number of variables required to simulate the atomistic structure of a perfectly ordered crystal. For solid solutions, this concept is commonly extended through the use of supercells, which are constructed by repeating the primitive or unit cell of the underlying lattice to form a larger simulation cell containing many more lattice sites. As the supercell size increases, both the computational cost of evaluating the energy of each configuration and the number of distinct atomic configurations grow rapidly, rendering exhaustive enumeration computationally prohibitive even for relatively simple energy models. While real materials are not constrained by supercell size, any computational results must nevertheless be assessed for convergence with respect to this approximation. 

Several classical approaches have been developed to find low-energy configurations of solid solutions, including Monte Carlo sampling, simulated annealing, and genetic algorithms, among others \cite{Purton2007,Mohn2015, Mohn2018, Allan2018}. 
In recent years, quantum technologies, and quantum annealing in particular, have attracted increasing attention as an alternative approach~\cite{Gusev2023,Choubisa2023,Binninger2025}. 
In our previous work~\cite{Camino2025}, we demonstrated how the investigation of solid solutions can be formulated as a Quadratic Unconstrained Binary Optimisation (QUBO)~\cite{Kochenberger2025,Lucas2014} model and implemented on D-Wave quantum annealers~\cite{Dwave}. 
We also showed how such devices can be used to recover thermodynamic properties of materials within the grand canonical ensemble.

In the present work, we take nitrogen-doped graphene as a test case to illustrate how a configurational optimisation problem can be mapped onto a Rydberg Hamiltonian~\cite{Browaeys2020,Henriet2020,Bluvstein2022} and implemented on neutral-atom quantum hardware, as realised on the Aquila QuEra device~\cite{Wintersperger2023,Wurtz2023}. 
Neutral-atom platforms offer several notable advantages: they operate at room temperature without the need for cryogenic cooling, allow for programmable atom positions and interaction ranges, and can function in both continuous-time (annealing) and digital modes. 
Recent experiments have demonstrated the possibility of dynamically moving atoms within the array, opening perspectives for highly reconfigurable quantum simulations~\cite{Bluvstein2022,Zhao2025}. 
Here, however, we focus exclusively on the annealing mode of operation.

The energy model for the material is mapped to the Rydberg Hamiltonian, which depends on only two tunable parameters: an on-site term corresponding to a controllable laser detuning, and a pair term representing the Rydberg-state interaction $C_6/R_{i,j}^6$, where $C_6$ is the interaction parameter between the atoms in the hardware and $R_{i,j}$ is the distance between sites $i$ and $j$ on the device. 
This mapping introduces substantial challenges beyond those encountered in our previous work~\cite{Camino2025}. 
The currently available hardware used in this work - the QuEra Aquila device - allows only two-dimensional atomic arrangements, motivating the choice of graphene as a model system for this proof of concept. 
Even within two dimensions, atom placement is subject to geometric constraints, such as minimum distances between sites and between adjacent rows. 
Furthermore, the energy window accessible through laser detuning is significantly narrower than the range of formation energies typically encountered in materials systems, requiring additional theoretical treatment.

{\color{black}Recent work has explored the statistical properties of quantum optimisation devices and programmable Rydberg systems, particularly in relation to whether their output distributions can be interpreted in thermodynamic terms. Experiments on large-scale Rydberg arrays have shown that the observed distributions are often better described in terms of coherent unitary dynamics rather than equilibrium thermal ensembles~\cite{Scholl2021, Ebadi2022}.}

{\color{black}At the same time, Boltzmann-like behaviour has been observed in several quantum optimisation approaches, where output probabilities scale approximately exponentially with energy, although this is usually established empirically and not linked to a thermodynamic interpretation~\cite{Lotshaw2023,Wurtz2024}. Hybrid quantum–classical methods also use quantum devices to generate low-energy configurations without assigning a statistical interpretation to the resulting distributions~\cite{Jeong2025}.}

{\color{black}In this work, we construct a mapping between a materials model and the Rydberg Hamiltonian under which the sampled configurations can be interpreted in thermodynamic terms. Our goal is to develop this mapping, as a proof of concept, and to assign a thermodynamic meaning to the hardware control parameters. A central component of the approach is the introduction of a rescaling factor, $\alpha_v$, which enables the sampled distributions to be described in terms of an effective temperature $T' = \alpha_v T$~\cite{Scholl2021}.}

We validate the proposed mapping by studying the nitrogen distribution in a 28-site graphene nanoflake, comparing the quantum hardware results with exhaustive classical enumeration across detuning (chemical potential) values. 
We then extend the analysis to a 78-site graphene nanoflake, for which full enumeration is computationally unfeasible, and benchmark the device output against {\color{black}random} Monte Carlo simulations.
Finally, we demonstrate that varying the interatomic separation allows one to sample effective distributions corresponding to different temperatures, confirming the thermodynamic consistency of the proposed scaling.

Overall, this work provides a proof-of-concept demonstration of the mapping between Density Functional Theory (DFT)~\cite{Hohenberg1964,Kohn1965} derived formation energies and a Rydberg Hamiltonian, along with a rescaling procedure that enables the implementation of realistic material models on available quantum hardware. 
Although currently limited to two-dimensional systems, the method is readily applicable to three-dimensional materials as hardware capabilities evolve~\cite{Barredo2018}. 
Because neutral atoms are also promising candidates for digital quantum computing, the approach developed here represents a potential building block for future hybrid workflows that combine continuous-time annealing and digital quantum algorithms for materials discovery.

\section*{Results}
\label{sec:results}
\begin{figure}
    \centering
    \includegraphics[width=0.99\linewidth]{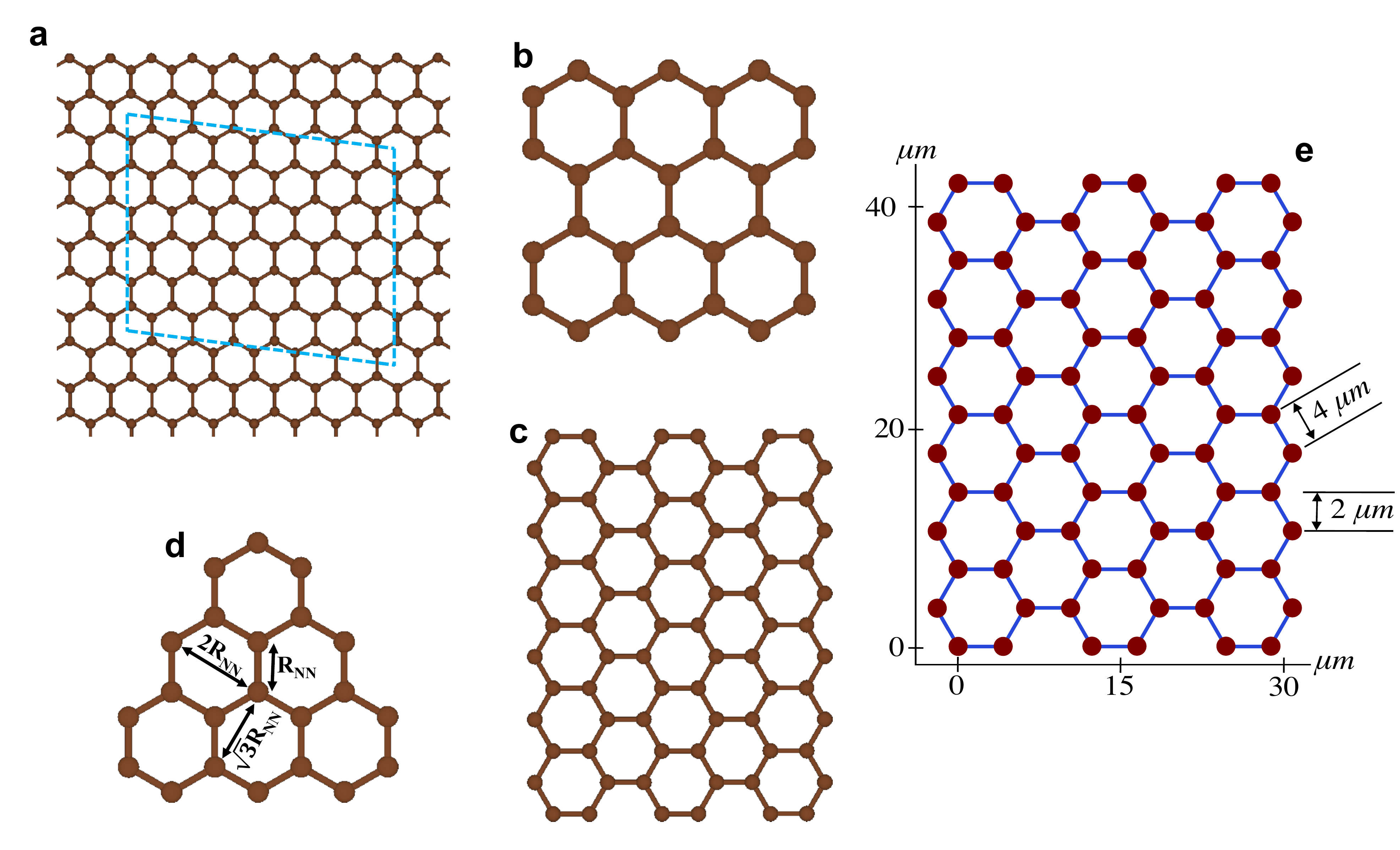}
    \caption{Structure of the graphene model used in this study and its mapping to the quantum hardware. 
\textbf{a} Periodic structure employed for the CRYSTAL DFT calculations. The dashed blue line indicates the supercell employed.
\textbf{b} Graphene nanoflake with 28 atoms used in the exhaustive search simulations. 
\textbf{c} Graphene nanoflake with 78 atoms used in the Monte Carlo simulations. 
\textbf{d} Zoom-in illustrating the distance between lattice sites in terms of the distance between nearest neighbours $R_{\mathrm{NN}}$. 
\textbf{e} Mapping of the structure in panel~\textbf{c} onto neutral atoms using $R_{\mathrm{NN}} = 4.0\,\mu$m. The minimum distance between neighbouring atoms (4$\mu$m) and neighbouring rows (2$\mu$m) allowed by the current hardware in the tight geometry is highlighted.
}
    \label{fig:graphene_model}
\end{figure}

\subsection*{Constructing the Rydberg Hamiltonian from DFT Data}
\label{sec: Constructing the Rydberg Hamiltonian from DFT Data}
We build the energy model used in the annealing from periodic Density Functional Theory (DFT) calculations. The computational details of these calculations can be found in the Methods section.

To be able to compare the energy of structures characterised by different nitrogen concentrations, we calculate the formation energy as:
\begin{equation}\label{formation_energy_1}
\Delta E_{k}^{f} =  E_{k} - \sum_{\chi^{i}} N_{k}^{\chi^{i}}E_{\chi^{i}},
\end{equation}
where $E_{k}$ is the total relaxed DFT energy per simulation cell for configuration $k$; $N_{k}^{\chi^{i}}$ is the number of $\chi^{i}$ atoms in configuration $k$; and $E_{\chi^{i}}$ is the DFT energy of species $\chi^{i}$ in its reference state. We used N$_2$ for the nitrogen energy reference and the energy of a carbon atom in the pristine graphene structure as a reference for carbon. This ensures that the $\Delta E_{k}^{f}$ for pristine graphene is zero and the energy of all doped structures is positive.

The 78-atom unit cell used in the DFT calculations is depicted in blue in panel \textbf{a} of Fig.~\ref{fig:graphene_model}. {\color{black}For each nitrogen concentration (1--10 atoms per cell), we generate a set of symmetry-independent configurations (SICs) by random substitution, discarding any configuration related to an existing one by the symmetry operations of the pristine graphene supercell. DFT geometry optimisations are performed only on the SICs; all symmetry-equivalent configurations (SECs) are generated afterwards by applying the space-group operations and inherit the energy of their parent SIC.  In this way, 28 SICs in the training set and 28 in the test set give rise to 2145 and 2184 SECs, respectively. The distribution of SICs across nitrogen concentrations is reported in Table~S1 of the Supplementary Information.} Their energies are calculated using the PBE~\cite{Perdew1996} functional and a pob-pVTZ~\cite{Vilela2019} basis set, as implemented in the CRYSTAL23 code~\cite{crystal23}. {\color{black}This regression-based construction differs from extracting interaction parameters from isolated one and two-impurity configurations, as it results in an effective interaction model averaged over the chemical environments sampled in the concentration range of interest.}

In a system of Rydberg atoms, the time evolution is governed by the following Hamiltonian \cite{Wurtz2023}:
\begin{equation}\label{eq:state_evolution_1}
    H(t) = \sum_{i=1}^N H_{\text{drive}, i}(t)  
    + \sum_{i=1}^{N-1}\sum_{j>i}^N H_{\text{vdW}, i, j},
\end{equation}
where
\begin{equation}\label{eq:state_evolution_2}
    H_{\text{drive}, i}(t) = \frac{\Omega(t)}{2}\left(e^{\mathrm{i}\phi(t)}|g_i\rangle\langle r_i| + e^{-\mathrm{i}\phi(t)}|r_i\rangle\langle g_i|\right) - \Delta_\text{g}(t)n_i,
\end{equation}
and 
\begin{equation}\label{eq:state_evolution_3}
    H_{\text{vdW}, i,j} =V_{i,j} \,n_i\, n_j = \frac{C_6}{R_{i,j}^6} \,n_i\, n_j.
\end{equation}
The term $H_{\text{drive}, i}(t)$ denotes a global Rabi drive that addresses the atoms simultaneously and uniformly, where $\Omega(t)$, $\phi(t)$ and $\Delta_g(t)$ denote the amplitude, phase and detuning of the driving field. Here $|g_i\rangle$ and $|r_i\rangle$ denote the ground and Rydberg states of the $i$-th atom and $n_i\equiv |r_i\rangle\langle r_i|$ is the occupation number operator.
{\color{black}This protocol implements quantum annealing by evolving the system from a regime in which the drive term dominates, enabling exploration of configurations, to a regime in which the interaction term $H_{\mathrm{vdW}}$ dominates and encodes the target energy landscape.}
At the beginning of the annealing protocol, all atoms are prepared in their ground states, which yields the many-body ground state of $H(t)$ for $\Omega(t)=0$, $\phi(t)=0$, and $\Delta_g(t)<0$. 
During the evolution process, we keep $\phi(t)=0$ and ramp $\Omega(t)$ (with a trapezoidal profile) while increasing $\Delta_g(t)$, driving the atoms into  superposition of the ground and Rydberg states.
At the end of the annealing process, we turn $\Omega(t)$ down to zero and set the global detuning to a final value $\Delta_g$, then perform a projective measurement, yielding the Rydberg occupation of each atom: $n_i=1$ if the $i$-th atom is in the Rydberg state otherwise $n_i=0$.
The term $H_{\text{vdW}, i,j}$ describes the two-body interaction between a pair of atoms, which depends on their occupation numbers $n_i$, separation $R_{i,j}$, and a hardware-specific constant $C_6$ (which is set to $5.42\times 10^{-24}\text{ rad m}^6/\text{s}$ in this work). We note that the interatomic separations $R_{i,j}$ are fixed throughout the annealing process.

The tunable parameters that allow the Rydberg Hamiltonian to encode a given problem are therefore the final global detuning $\Delta_\text{g}$ and the interatomic separations~$R_{i,j}$. The formation energies of the structures in the training set are mapped to the Rydberg Hamiltonian by minimising the following linear regression:
\begin{equation}
\label{eq:linear_regression_ryd}
    \min_{\{V^{\mathrm{DFT}},\, V_{i,j}^{\mathrm{DFT}}\}}
    \sum_{k=1}^{N_{\text{train}}} 
    \left[
        \Delta E_{k}^{f} - 
        \left(
            \sum_{i} V^{\mathrm{DFT}} n_{i}^{k}
            + 
            \sum_{i j} V_{i,j}^{\mathrm{DFT}} n_{i}^{k} n_{j}^{k}
        \right)
    \right]^{2},
\end{equation}
where $N_{\text{train}}$ is the number of structures in the training set, $\Delta E_{k}^{f}$ is the formation energy for structure $k$ as defined in Eq.~\ref{formation_energy_1}, $V^{\mathrm{DFT}}$ is the on-site energy, which is the same for all sites, and $V_{i,j}^{\mathrm{DFT}}$ is the two-body potential between sites $i$ and $j$. The variables $n_{i}^{k}$ are the components of the site-occupation vector $\mathbf{n}^{k}$ for configuration $k$. Here, $n_{i}^{k} = 1$ implies that site $i$ is occupied by a nitrogen atom, and $n_{i}^{k} = 0$ a carbon atom.
By writing the two-body potential explicitly, we obtain:
\begin{equation}\label{eq:mapping_1}
        \min_{\{V^{\mathrm{DFT}},\, R_{i,j}^{\mathrm{DFT}}\}}\sum_{k=1}^{N_{\text{train}}} \left( \Delta E_{k}^{f} - \left(\sum_{i} V^{\mathrm{DFT}} n_{i}^{k} + \sum_{ij} \frac{C_6}{(R_{i,j}^{\mathrm{DFT}})^6} n_{i}^{k} n_{j}^{k} \right) \right)^{2},
\end{equation}
where the two-body potential has been explicitely expressed through its dependence on the distance between sites $R_{i,j}^{\mathrm{DFT}}$. {\color{black}We emphasise that the $1/r^6$ dependence is introduced here because it matches the native interaction of the Rydberg platform; it should therefore be understood as an effective, hardware-constrained approximation to the DFT energetics rather than as a first-principles model of the nitrogen--nitrogen interaction itself. Its suitability is assessed \textit{a posteriori} through the quality of the regression on the DFT data.}  

We can simplify Eq.~\ref{eq:mapping_1} further by taking into consideration two factors. First, in the unit cell displayed in panel \textbf{a} of Fig.~\ref{fig:graphene_model}, all sites are equivalent. This means that the $V_{i,j}$ term is the same for all for all nearest-neighbour pairs $i$ and $j$. Therefore, we can define a single distance between neighbouring sites $R_{\mathrm{NN}}$ that applies to all couples $i$ and $j$ of nearest neighbours. The distances between next nearest neighbours can be obtained by using simple geometrical considerations as shown in the panel \textbf{d} of Fig.~\ref{fig:graphene_model}. Secondly, because $H_{\text{vdW}, i, j}$ scales as ${R_{i,j}^{-6}}$, the mapping can be truncated at the fourth nearest neighbours, for which the interaction energy has already decayed to $\tfrac{1}{343}V_{\mathrm{NN}}$ (the fourth nearest neighbour in a heaxagonal lattice is at a distance of $\sqrt{7}R_{\mathrm{NN}}$). Therefore, Eq.~\ref{eq:mapping_1} becomes:
\begin{align}\label{eq:linear_regression_ryd_simple}
& \min_{\{V^{\mathrm{DFT}},\, R_{\mathrm{NN}}^{\mathrm{DFT}}\}} \sum_{k=1}^{N_{\text{train}}} \bigg( 
    \Delta E_{k}^{f} - \bigg(
    \sum_{i} V^{\mathrm{DFT}} n_{i}^{k} + \nonumber \\
    & \sum_{ij \in \mathrm{NN}} \frac{C_6}{(R_{\mathrm{NN}}^{\mathrm{DFT}})^6} n_{i}^{k} n_{j}^{k} + 
     \sum_{ij \in \text{NNN}} \frac{C_6}{\left(\sqrt{3} R_{\mathrm{NN}}^{\mathrm{DFT}}\right)^6} n_{i}^{k} n_{j}^{k} + \nonumber \\
    & \sum_{ij \in \text{3NN}} \frac{C_6}{\left(2 R_{\mathrm{NN}}^{\mathrm{DFT}}\right)^6} n_{i}^{k} n_{j}^{k} + 
     \sum_{ij \in \text{4NN}} \frac{C_6}{\left(\sqrt{7} R_{\mathrm{NN}}^{\mathrm{DFT}}\right)^6} n_{i}^{k} n_{j}^{k}
    \bigg)
\bigg)^2,
\end{align}
where the sums were limited to the fourth nearest neighbours (4NN). From Eq.~\ref{eq:linear_regression_ryd_simple} we obtain the $V^{\mathrm{DFT}}$ and $R_{\mathrm{NN}}^{\mathrm{DFT}}$ that will be used in the mapping to the quantum annealer.

Using Eq.~\ref{eq:linear_regression_ryd_simple} we obtain $V^{\mathrm{DFT}} = 3.613 \times 10^{-4}$ eV and $R_{\mathrm{NN}}^{\mathrm{DFT}} = 1.6122\mu$m. The model yields a correlation coefficient of $R = 0.962$ on the training set, a mean squared error of $3.13  \times 10^{-4}$ eV on the test set, and a Spearman’s rank correlation coefficient of $\rho = 0.986$. {\color{black}The convergence of the regression with respect to the number of training SICs and the choice of the fourth-nearest-neighbour truncation are analysed in Figures~S1 and~S2 of the Supplementary Information.}

\subsection*{Setting the Composition through Detuning and Chemical Potential}
When computing the formation energies $\Delta  E_{k}^{f}$ from Eq.~\ref{formation_energy_1}, pure graphene has $\Delta  E_{k}^{f}=0$ and all structures containing nitrogen atoms have $\Delta  E_{k}^{f}>0$.
As quantum annealing typically finds the ground state solution, using the $V^{\mathrm{DFT}}$ and $R_{\mathrm{NN}}^{\mathrm{DFT}}$ values derived from Eq.~\ref{eq:linear_regression_ryd_simple} would result in identifying only the trivial solution consisting entirely of carbon atoms.

In our previous work~\cite{Camino2025}, we mapped this class of problems onto QUBO models that were solved on D-Wave superconducting qubit quantum annealers and demonstrated that the outcome can be tuned toward a targeted composition by introducing a penalty on the on-site term of the QUBO expression.
There, we further established the connection between this linear bias and the chemical potentials of the species in reservoirs in equilibrium with the material. {\color{black}Experimentally, a species-dependent chemical potential provides a way to control the relative concentration of different atomic species, corresponding in practice to varying growth conditions or chemical environments. In our mapping, this is implemented via the global detuning~\cite{Zhang1991}.}
To make this correspondence explicit, we introduce the grand-canonical formation energies as
\begin{equation}\label{eq:formation_energy_gc}
    \Delta \tilde{E}_{k}^{f}
    = \Delta E_{k}^{f}
      + \sum_{\chi^{i}} N_{k}^{\chi^{i}} \mu_{\chi^{i}},
\end{equation}
where $\Delta E_{k}^{f}$ is the canonical (fixed-composition) formation energy
defined in Eq.~\ref{formation_energy_1}, and $N_{k}^{\chi^{i}}$ and
$\mu_{\chi^{i}}$ denote, respectively, the number of atoms of species
$\chi^{i}$ in configuration $k$ and the corresponding chemical potential.

We define the chemical potential difference
$\Delta\mu = \mu_{\mathrm{N}} - \mu_{\mathrm{C}}$ and write
$\mu_{\mathrm{N}} = \mu_{\mathrm{C}} + \Delta\mu$, which for our chosen system gives
\begin{equation}
\Delta \tilde{E}_{k}^{f}
    = \Delta E_{k}^{f}
      + (N_k^{\mathrm{C}} + N_k^{\mathrm{N}}) \mu_{\mathrm{C}}
      + N_k^{\mathrm{N}} \Delta\mu.
\end{equation}
Since we want $\Delta \tilde{E}_{k}^{f}= \Delta E_{k}^{f}$ when there are no nitrogen atoms available, we set $\mu_C$ = 0, yielding
\begin{equation}
\Delta \tilde{E}_{k}^{f}
    = \Delta E_{k}^{f} + N_k^{\mathrm{N}} \Delta\mu.
\end{equation}
In the following sections, we use this form, in which the thermodynamics is
fully controlled by the single parameter $\Delta\mu$.

A key consequence of this expression is that the chemical potential term adds
a contribution that is linear in the number of nitrogen atoms
$N_k^{\mathrm{N}}$. In our encoding, nitrogen corresponds to $n_i = 1$ in the
binary configuration returned by the quantum annealer. Therefore, introducing
$\Delta\mu$ does not require refitting the parameters
$V^{\mathrm{DFT}}$ and $R_{\mathrm{NN}}^{\mathrm{DFT}}$ obtained from
Eq.~\ref{eq:linear_regression_ryd_simple}. Instead, the chemical potential can
be incorporated directly into the mapping as a shift of the on-site term:
\begin{equation}\label{eq:mapping_2}
    \Delta \tilde{E}_k^{f}
    \approx
    \sum_{i} \left( V^{\mathrm{DFT}} + \Delta\mu \right) n_{i}^{k}
    + \sum_{i j}
        \frac{C_6}{(R^{\mathrm{DFT}}_{i,j})^6}
        n_{i}^{k} n_{j}^{k},
\end{equation}
where $\approx$ indicates that this expression represents the optimal Rydberg-Hamiltonian approximation to the DFT formation energies achievable
under the hardware constraints.

By comparing Eqs.~\ref{eq:state_evolution_1} and \ref{eq:mapping_2}, we find that the detuning parameter $\Delta_g$ in the quantum annealer corresponds to the term $(V^{\mathrm{DFT}} + \Delta\mu)$ in our energy model. This correspondence forms the basis for the subsequent analysis, establishing a direct physical link between the thermodynamic control of chemical potential in the grand canonical ensemble and the tunable laser detuning $\Delta_g$ in the quantum hardware.

\subsection*{Implementing the Mapping within Hardware Constraints}
\label{subsec:hardware_constraints}
When mapping the problem to a set of Rydberg atoms, as implemented in the QuEra Aquila device using the experimental setup, we need to take into account the following constraints:
\begin{itemize}
    \item the maximum number of atoms is 256
    \item the maximum size of the atomic arrangement is  $76\mu\text{m}\times 75\mu\text{m}$
    \item the range of global detuning $\Delta_g$ is $[-\Delta_g^{\mathrm{max}}, \Delta_g^{\mathrm{max}}]$ where $\Delta_g^{\mathrm{max}}=8.227649\times10^{-8}$eV
    \item the minimum distance allowed between two atoms is $R_{\mathrm{NN}}^{\mathrm{min}}=4\mu\text{m}$
    \item the minimum distance allowed between two rows of atoms in 2$\mu\text{m}$ (this is available in the tight geometry setup; the default is $4\mu\text{m}$)
\end{itemize}
The latter two constraints are depicted in panel \textbf{e} of Fig~\ref{fig:graphene_model}.

\begin{figure}
    \centering    \includegraphics[width=0.99\linewidth]{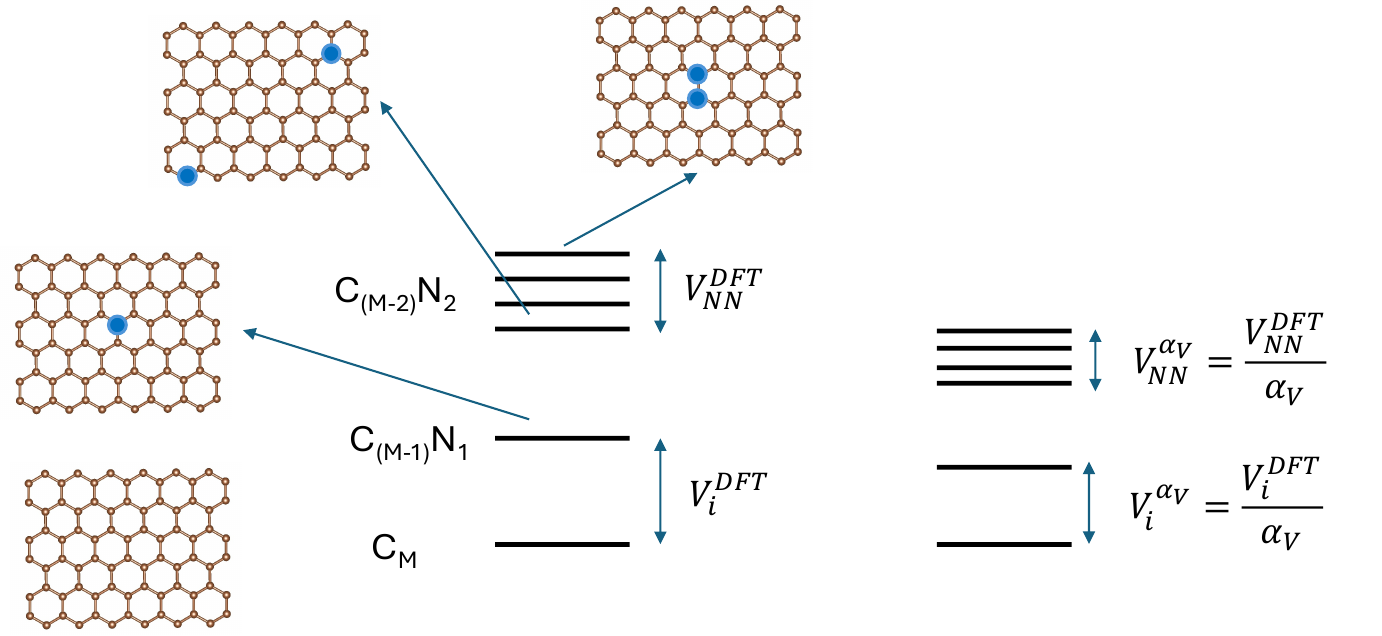}
    \caption{Effect of scaling on the energy levels in the material and the hardware. Energy levels of pure graphene (C$_{\text{M}}$) and graphene doped with one (C$_{(\text{M}-1)}$N) or two (C$_{(\text{M}-2)}$N$_2$) nitrogen atoms. $V^{\mathrm{DFT}}$ denotes the nitrogen on-site energy, and $V_{\mathrm{NN}}^{\mathrm{DFT}}$ the two-body potential between neighbouring sites. The terms $\alpha_v$ is the scaling factor used to map the DFT energies onto the hardware, subject to its constraints.  $V_{\mathrm{NN}}^{\alpha_v}$ is the scaled two-body potential between neighbouring sites, and $V^{\alpha_v}$ is the nitrogen on-site energy scaled consistently.  
    }
    \label{fig:energy_levels}
\end{figure}

Specifically, due to constraints on the accessible global detuning range and the minimum allowed interatomic spacing, we cannot directly realise on this hardware the DFT-derived parameters {\color{black}$V^{\mathrm{DFT}}_{\mathrm{NN}}$ because it would require the atoms in the hardware to be placed at $R_{\mathrm{NN}}^{\mathrm{DFT}} = 1.6122 \mu\text{m}$, which is approximatively $0.4\times R_{\mathrm{NN}}^{\mathrm{min}}$}.
To address this mismatch, we develop and apply a systematic rescaling procedure that maps the DFT parameters to experimentally feasible values. This procedure is essential for modelling realistic atomic systems and constitutes a central contribution of this work.
To illustrate the rescaling procedure, we begin with the energy levels of the nitrogen-doped graphene system, shown in Fig.~\ref{fig:energy_levels}. The left panel presents the progression of energy levels from pure graphene to configurations containing two nitrogen atoms.
The term $V^{\mathrm{DFT}}$ represents the on-site energy associated with introducing a single nitrogen atom into the structure. To simplify the derivation of the equations needed for rescaling the energies, we assume here that all sites are equivalent, so that all configurations with just one carbon atom replaced with one nitrogen atom are degenerate. 
For structures containing two nitrogen atoms separated by more than four neighbours, we assume negligible interaction between nitrogen atoms, so the total energy of the structures is simply $2V^{\mathrm{DFT}}$. In contrast, when the two nitrogen atoms occupy nearest-neighbour sites, the total energy becomes $2V^{\mathrm{DFT}} + V_{\mathrm{NN}}^{\mathrm{DFT}}$, where
\begin{align}
    V_{\mathrm{NN}}^{\mathrm{DFT}} \equiv \frac{C_6}{(R_{\mathrm{NN}}^{\mathrm{DFT}})^6}
\end{align}
is the two-body interaction energy between neighbouring sites as derived from Eq.~\ref{eq:linear_regression_ryd}. Therfore, the energy separation between the highest and lowest-energy two-nitrogen configurations is $V_{\mathrm{NN}}^{\mathrm{DFT}}$. The energies discussed so far are those obtained by fitting the DFT results to the Rydberg Hamiltonian,
\textit{i.e.}, solving Eq.~\ref{eq:linear_regression_ryd_simple}. 
However, the values of $V_{\mathrm{NN}}^{\mathrm{DFT}}$ lie beyond the hardware capabilities. In particular, the largest achievable two-body interaction between two nearest-neighbour atoms is
\begin{align}
    V_{\mathrm{NN}}^{\alpha_v}\equiv\frac{C_6}{(R_{\mathrm{NN}}^{\mathrm{min}})^6},
\end{align}
which is smaller than the DFT nearest-neighbour interaction $V_{\mathrm{NN}}^{\mathrm{DFT}}$ because $R_{\mathrm{NN}}^{\mathrm{DFT}} < R_{\mathrm{NN}}^{\mathrm{min}}$. {\color{black}This is depicted on the right hand side of Fig.~\ref{fig:energy_levels}}. To overcome this limitation, we introduce a rescaling factor
\begin{align}
\label{eq:def_alpha_v}
    \alpha_v \equiv \frac{V_{\mathrm{NN}}^{\mathrm{DFT}}}{V_{\mathrm{NN}}^{\alpha_v}} = \left(\frac{R_{\mathrm{NN}}^{\mathrm{min}}}{R_{\mathrm{NN}}^{\mathrm{DFT}}}\right)^6\approx236.69.
\end{align}
With this choice, reproducing the energy spectrum of the nitrogen-doped graphene system reduces to rescaling the Rydberg-atom energies by $\alpha_v$. {\color{black}We note that while similar rescaling strategies are routinely used in quantum simulation~\cite{Scholl2021}, here the rescaling plays a different role, which is to establish a correspondence between the distribution sampled by the hardware and the thermodynamic distribution of the underlying material model.} For consistency, we also rescale the on-site energy according to:
\begin{align}
\label{eq:def_rescaled_on_site_energy}
    V^{\alpha_v} \equiv \frac{V^{\mathrm{DFT}}}{\alpha_v}.
\end{align}
We emphasise that introducing $V^{\alpha_v}$ is not enforced by the hardware capability, unlike $V_{\mathrm{NN}}^{\alpha_v}$; however, the accessible range of the corresponding chemical potential remains limited by the detuning range achievable on the hardware. {\color{black}Rather, its introduction enforces a uniform rescaling of the Hamiltonian, ensuring that the distribution generated by the hardware can be interpreted as a Boltzmann-like distribution corresponding to that of the original (unscaled) system, but at a higher effective temperature. In the following, we show how this correspondence can be derived.}

First, we consider the partition function in the grand canonical ensemble for the nitrogen-doped graphene:
\begin{equation}\label{eq:pf_gce}
    \Xi = \sum_{k=0}^{N_n} M_k \exp{\frac{-\Delta \tilde{E}_{k}^{f}}{k_BT}}  = \sum_{k=0}^{N_n} M_k \exp{\frac{-(\Delta E_{k}^{f}+N_k^N\Delta\mu)}{k_BT}}, 
\end{equation}
where $T$ is the effective temperature to be determined in the exhaustive search section, $M_k$ is the multiplicity of state $k$ and all the other quantities have been defined above.
To derive the grand canonical partition function $\Xi$ under the hardware constraints, we begin by rescaling the exponent in Eq.~\ref{eq:pf_gce} by $\alpha_v$:

\begin{align}
    \Xi(T,\Delta\mu,\alpha_v) &= \sum_{k=0}^{N_n} M_k \exp\left(\frac{-\Delta E_{k}^{f} - N_k^{N} \Delta\mu}{k_B T \alpha_v}\right)
    \label{eq:pf_gce_alpha_a} \\      
    &= \sum_{k=0}^{N_n} M_k \exp\left[\frac{-1}{k_B T} \left( \frac{\sum_{i} (V^{\mathrm{DFT}} + \Delta\mu) n_{i}^{k} + \sum_{ij} V_{i,j}^{\mathrm{DFT}} n_{i}^{k} n_{j}^{k}}{\alpha_v} \right)\right]
    \label{eq:pf_gce_alpha_b} \\    
    &= \sum_{k=0}^{N_n} M_k \exp\left[\frac{-1}{k_B T} \left( \sum_{i}\frac{ (V^{\mathrm{DFT}} + \Delta\mu) }{\alpha_v}n_{i}^{k} + \sum_{ij}\frac{ C_6 }{(R_{i,j}^{\mathrm{hw}})^6} n_{i}^{k} n_{j}^{k}\right)\right]
    \label{eq:pf_gce_alpha_e} \\    
    &= \sum_{k=0}^{N_n} M_k \exp\left[\frac{-1}{k_B T} \left( -\sum_{i} \Delta_g n_{i}^{k} + \sum_{ij}\frac{ C_6 }{(R_{i,j}^{\mathrm{hw}})^6}n_{i}^{k} n_{j}^{k} \right)\right]
    \label{eq:pf_gce_alpha_f}
\end{align}
where we have used Eq.~\ref{eq:linear_regression_ryd} in the derivation. As explained above, $V_{\mathrm{NN}}^{\mathrm{DFT}}/\alpha_v$ is the largest achievable two-body interaction when the nearest-neighbour sites are separated by $R_{\mathrm{NN}}^{\mathrm{min}}$. {\color{black}We use $R_{i,j}^{\mathrm{hw}}$ to denote the distance between sites $i$ and $j$ in the hardware embedding, where nearest-neighbour sites are separated by $R_{\mathrm{NN}}^{\mathrm{min}}$.} Furthermore, the linear term in $n_i^k$ in the exponential in Eq.~\ref{eq:pf_gce_alpha_e} has the same structure as the global detuning term in the Rydberg Hamiltonian (see Eq.~\ref{eq:state_evolution_2}), we therefore obtain \ref{eq:pf_gce_alpha_f} by making the identification
\begin{align}
\label{eq:def_Delta_mu^alpha_v}
    \Delta_g = -\bigl(V^{\alpha_v} + \Delta\mu^{\alpha_v}\bigr),
    \qquad
    \Delta\mu^{\alpha_v} \equiv \frac{\Delta\mu}{\alpha_v}.
\end{align}
Here, $\Delta\mu^{\alpha_v}$ is the rescaled chemical potential, which plays a role analogous to $V^{\alpha_v}$ in enforcing a consistent overall energy rescaling.
By comparing Eq.~\ref{eq:pf_gce_alpha_f} and Eq.~\ref{eq:pf_gce_alpha_a}, we conclude that the partition function $\Xi$ sampled by the annealer at the effective sampling temperature $T$, global detuning $\Delta_g$, and minimum interatomic spacing $R_{i,j}^{\mathrm{hw}}$ corresponds to the partition function of the material evaluated at the rescaled temperature 
\begin{align}
    \label{eq:def_rescaled_temperature}
    T' = T \alpha_v,
\end{align}
where $\alpha_v$ is given in Eq.~\ref{eq:def_alpha_v}.

Eqs.~\ref{eq:def_alpha_v}, \ref{eq:def_rescaled_on_site_energy}, \ref{eq:def_Delta_mu^alpha_v} and \ref{eq:def_rescaled_temperature} define the mapping to implement the nitrogen-doped graphene model on the Rydberg-atom quantum hardware. 
Under this mapping, desirable energy scales, typically obtained from DFT calculations, that are inaccessible to the hardware become accessible due to a uniform rescaling, so that configurations that would be thermally populated in the physical system at temperature $T'$ correspond to those populated in the hardware at an effective temperature $T$.
This rescaling also modifies how we interpret the detuning. Per Eq.~\ref{eq:def_Delta_mu^alpha_v}, we can explore a chemical potential range for the nitrogen-doped graphene model by varying the detuning $\Delta_g\in[-\Delta_g^{\mathrm{max}}, \Delta_g^{\mathrm{max}}]$. Within this interval, $\Delta_g=0$ maps to $\Delta\mu=-V^{\mathrm{DFT}}$, as shown in the center of the bottom panel of Fig.~\ref{fig:energy_levels}. Because of the rescaling, the annealer effectively samples the chemical potential of the nitrogen-doped graphene model within the range $\Delta\mu \in [-V^{\mathrm{DFT}} - \alpha_v \Delta_g^{\mathrm{max}}, -V^{\mathrm{DFT}} + \alpha_v \Delta_g^{\mathrm{max}}]$. 
Since $\alpha_v\approx236.69$, the range of $\Delta\mu$ is two orders of magnitude larger than the energy scale of the hardware, as schematically illustrated in the bottom panel of Fig.~\ref{fig:energy_levels}.
Rescaling is a critical step in our mapping, and any detuning value applied on the hardware, $\Delta_g$, must therefore be multiplied by $\alpha_v$ to recover the corresponding effective chemical potential in the DFT-derived energies. In other words, the annealer samples the distribution associated with $\Delta\mu^{\alpha_v}$, not the original $\Delta\mu$. This distinction is crucial when comparing thermodynamic averages to classical calculations obtained, for example, from an exhaustive search or a Monte Carlo approach, or when benchmarking against experimental data.

\subsection*{Benchmarking the Quantum Annealing against an Exhaustive Search}
\label{subsec:es}
\begin{figure}[!]
    \centering    \includegraphics[width=0.99\linewidth]{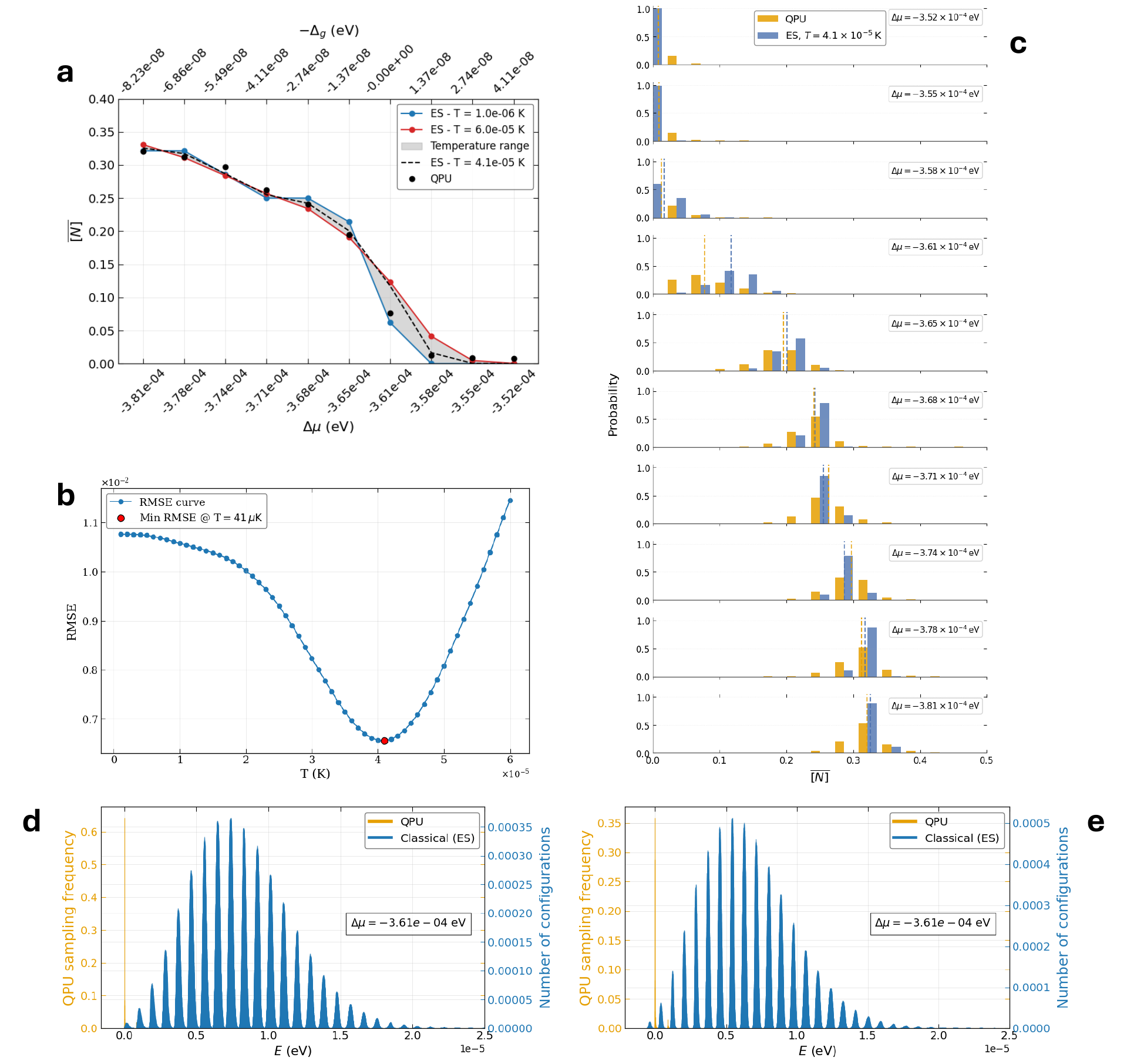}
    \caption{QPU and exhaustive search data for the 28-site structure shown in Fig.~\ref{fig:graphene_model}\textbf{b}. Panel \textbf{a} reports the average nitrogen concentration as a function of $\Delta\mu$ (bottom $x$-axis) and $\Delta_g$ (top $x$-axis). Black dots indicate QPU data obtained at $R_{\mathrm{NN}} = 4 \mu$m. Blue and red curves correspond to exhaustive search results at $T = 1 \mu$K and $T = 60 \mu$K, respectively, with the grey shaded area marking the range between these two temperatures. The dashed black line shows the result at $T = 41 \mu$K, which minimises the RMSE with respect to the QPU data, as highlighted by the red dot in panel \textbf{b}. Each subplot in panel \textbf{c} corresponds to a different value of the chemical potential and displays the distribution of nitrogen concentration. The orange and blue dashed vertical lines mark the average concentration obtained from the QPU and from the exhaustive search at $T = 41 \mu$K, respectively. Solid bars of the same colours illustrate the spread of concentrations around these mean values. Panels \textbf{d} and \textbf{e} show histograms of the energies of the configurations sampled by the QPU (orange) and obtained from the exhaustive search (blue) for $\Delta\mu=-3.81\times10^{-4}$ eV and $\Delta\mu=-3.61\times10^{-4}$ eV, respectively. The exhaustive-search curve corresponds to the density of states (number of configurations at a given energy), while the QPU curve reflects the frequency with which configurations of a given energy are sampled during the annealing process. These are not Boltzmann-weighted probability distributions.
    }
    \label{fig:exhaustive_search}
\end{figure}

In this section, we benchmark the mapping of the doped graphene system to the Rydberg atom hardware introduced above. Specifically, we estimate the effective sampling temperature of the Rydberg-atom system, a free parameter in the mapping, by running a small instance on the QPU and comparing the resulting average nitrogen concentration against that obtained classically through an exhaustive search.
For this benchmark, we use the 28-site system depicted in Fig.~\ref{fig:graphene_model} panel \textbf{b} and map it to a Rydberg-atom geometry with nearest-neighbour spacing fixed to the hardware minimum $R_{\mathrm{NN}}^{\mathrm{min}}=4\mu\text{m}$. We then execute the annealing protocol at ten values of $\Delta_g$ in $[-\frac{1}{2}\Delta_g^{\mathrm{max}},\Delta_g^{\mathrm{max}}]$. The details of the annealing schedules are reported in the Methods section. At the end of the annealing, we measure the Rydberg occupation and calculate the average concentration of nitrogen atoms as:
\begin{equation}
    \overline{[N]}^{\mathrm{qa}}(\Delta_g)
    = \frac{1}{N_{qa}}
      \sum_{s=1}^{N_{qa}}
      \|\mathbf{n}_s\|_{1},
\end{equation}
where $\overline{[N]}^{\mathrm{qa}}(\Delta_g)$ denotes the average nitrogen concentration obtained at the value of $\Delta_g$, $N_{qa}$ is the number of times the quantum annealing was performed, $\mathbf{n}_s$ is the binary vector returned by the QPU for sample $s$, and $\|\mathbf{n}_s\|_1$ denotes its $\ell^1-norm$ (\textit{i.e.}, the Hamming weight, or the number of ones in the vector). Since nitrogen sites are mapped to 1 and $\Delta_g$ maps to the chemical potential of the doped graphene, we interpret $\overline{[N]}^{\mathrm{qa}}(\Delta_g)$ as the average nitrogen concentration, which is a function of the chemical potential, obtained from the QPU. These are shown as black dots, in panel~\textbf{a} of Fig.~\ref{fig:exhaustive_search}. 

As described in the mapping to hardware, the QPU-estimated average nitrogen concentration corresponds to that of the doped graphene at a certain rescaled temperature. To determine that, we compare $\overline{[N]}^{\mathrm{qa}}(\Delta_g)$ against the exact equilibrium prediction from the exhaustive search over the 28-site model across the same set of chemical potentials, and fit for the temperature that best matches the QPU data.
Specifically, we calculate the average composition by performing an exhaustive search on the whole set of possible configurations containing between zero and twenty-eight nitrogen atoms. In this simulation, the energy of each configuration is calculated as:
\begin{equation}\label{eq:energies_es_1}
    E^{\mathrm{Ry}}_k = \sum_i \Delta_g n_{i}^{k} + \sum_{i>j} \frac{C_6}{R_{i,j}^{\mathrm{min}}}n_{i}^{k} n_{j}^{k},
\end{equation}
which is the Rydberg energy (see Eq.~\ref{eq:state_evolution_1} and Eq.~\ref{eq:state_evolution_2}) at the end of the annealing. 
The average nitrogen composition $\overline{[N]}^{\mathrm{es}}(\Delta_g)$ obtained from the exhaustive search at global detuning $\Delta_g$ is determined by using:
\begin{equation}\label{eq:average_conc_hardware}
    \overline{[N]}^{\mathrm{es}}(\Delta_g)
    = \sum_{k} p_k\, \|\mathbf{n}_k\|_{1},
\end{equation}
where $\mathbf{n}_k$ is the binary vector mapping of structure $k$ and the probability $p_k$ of observing structure $k$ is calculated classically as:
\begin{equation}\label{eq:probabilities_classical}
    p_k = \frac{\exp\frac{-E^{\mathrm{Ry}}_k}{k_BT}}{\sum_k \exp\frac{-E^{\mathrm{Ry}}_k}{k_BT}}.
\end{equation}
Here $T$ is the effective sampling temperature of the annealer (see Eq.~\ref{eq:def_rescaled_temperature}). Inspired by the hardware temperature being in the micro Kelvin regime, we sampled sixty temperatures from $1\times10^{-6}$K to $6\times10^{-5}$K. The resulting values are reported in Fig.~\ref{fig:exhaustive_search} panel \textbf{a}. Here we plot the $\overline{[N]}^{\mathrm{es}}(\Delta_g)$ for the minimum (blue line) and maximum (red line) temperatures sampled and shade the interval between the two (grey area). 
To identify the effective sampling temperature, we calculated the root mean squared error (RMSE) between each of the curves obtained at the temperatures we sampled and the QPU data (Fig.~\ref{fig:exhaustive_search} panel \textbf{b}). We find the best fit (RMSE=6.55$\times 10^{-3}$) to be at $T$=41 $\mu$K, 
and the $\overline{[N]}^{\mathrm{es}}$ at this temperature is plotted as a dashed line in Fig.~\ref{fig:exhaustive_search}\textbf{a}.
After extracting the annealer’s effective sampling temperature from the average nitrogen concentration, we further validate the mapping by comparing the full distribution of the nitrogen concentration measured on the QPU with the corresponding classical predictions. 
For this, we calculate the energy of the complete set of configurations by using:
\begin{equation}\label{eq:energies_es_2}
    E^{\mathrm{DFT}}_k = \sum_{i} (V^{\mathrm{DFT}} + \Delta\mu) n_{i}^{k} + \sum_{i>j} V_{i,j}^{\mathrm{DFT}} n_{i}^{k} n_{j}^{k}.
\end{equation}
From Eq.~\ref{eq:def_Delta_mu^alpha_v}, the values of $\Delta\mu$ sampled range between $-V^{\mathrm{DFT}}-\alpha_v \Delta_g^{\mathrm{max}}$ and $-V^{\mathrm{DFT}}+\alpha_v \Delta_g^{\mathrm{max}}$. 
The average composition is then calculated using Eqs.~\ref{eq:average_conc_hardware} and \ref{eq:probabilities_classical} with $E^{\mathrm{Ry}}_k$ replaced by $E^{\mathrm{DFT}}_k$ and $T$ replaced by the scaled temperature $T' = \alpha_v T$ defined in Eq.~\ref{eq:def_rescaled_temperature}.
{\color{black}In Fig.~\ref{fig:exhaustive_search}\textbf{c}, each panel corresponds to a different value of $\Delta\mu$. The orange and blue bars show the nitrogen concentration spread around the average (dotted line) for the QPU and exhaustive search data at the rescaled temperature $T' = \alpha_v T = 9.71$ mK ($T=41 \mu K$), respectively. We obtain a Total Variation Distance (TVD) of 0.294 between the QPU and classical distributions. The TVD takes values between 0 (identical distributions) and 1 (completely disjoint distributions), indicating a moderate level of agreement. Despite this deviation, both the mean occupation and the spread of the distributions are well reproduced, showing that the QPU captures the main thermodynamic features of the system.}
{\color{black}The largest deviation occurs at $\Delta\mu=-3.61\times10^{-4}$ eV ($\Delta_g = 0$). At this detuning, the linear term does not favour nitrogen incorporation, so the energies are determined primarily by the pair-interaction term. As a result, many configurations lie very close in energy, including a large number of low-energy and zero-energy states. In this regime, the Boltzmann weights become more sensitive to temperature, since small energy differences between low-lying configurations lead to larger relative changes in their statistical weights. This increased temperature sensitivity makes the deviation between the QPU and exhaustive-search distributions more pronounced.}

{\color{black}An important result is displayed in panels \textbf{d} and \textbf{e} of Fig.~\ref{fig:exhaustive_search}, where we show histograms of the energies of the configurations explored in the exhaustive search (blue line) and those sampled by the QPU (orange line) at $\Delta\mu=-3.81\times10^{-4}$~eV ($\Delta_g=-\Delta_g^{\mathrm{max}}$) and $\Delta\mu=-3.61\times10^{-4}$~eV ($\Delta_g=0$), respectively. 
The exhaustive-search curve shows to the number of configurations at each energy (\textit{i.e.}, the density of states), while the QPU curve displays the frequency with which configurations of a given energy are sampled during the annealing process.
As the exhaustive search enumerates all configurations by construction, its distribution spans the full energy range of the system. In contrast, the QPU samples only the lowest-energy portion of the spectrum, reflecting its tendency to explore configurations close to the global minimum. For $\Delta\mu=-3.81\times10^{-4}$~eV, the majority of the states returned by the annealer have negative energies, confirming that the device predominantly identifies configurations that contain nitrogen atoms arranged to minimise pair interactions. At $\Delta\mu=-3.61\times10^{-4}$~eV, the lowest accessible energy is zero, and the sampled states indeed cluster around this value.
In both panels, the left and right vertical axes use markedly different scales. This choice highlights that, while the exhaustive search samples the full energy spectrum, the QPU predominantly explores the low-energy sector of the landscape.}

\subsection*{Monte Carlo Benchmarking of the Quantum Sampling}
\label{subsec:mc}
\begin{figure}
    \centering    \includegraphics[width=0.99\linewidth]{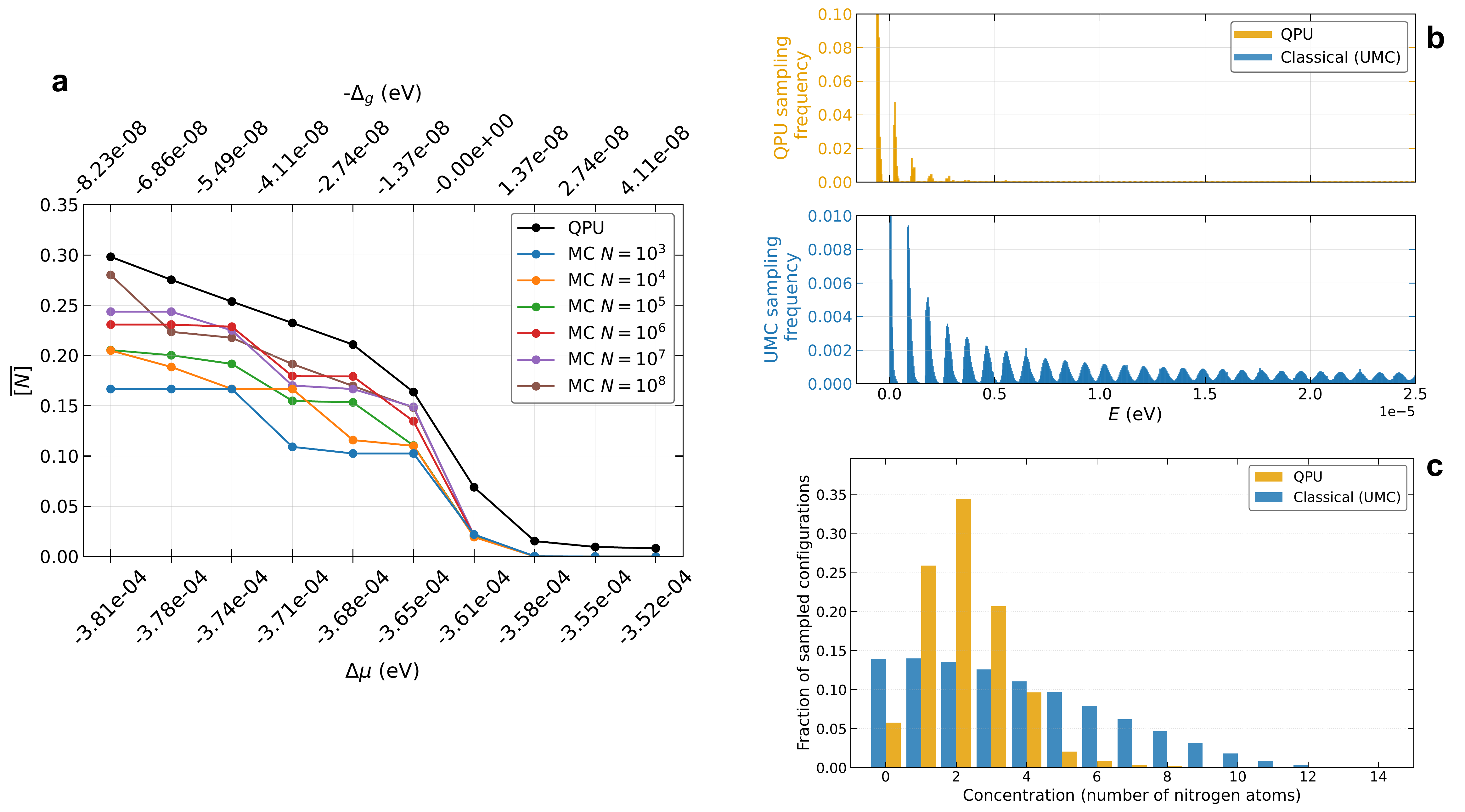}
    \caption{{\color{black}QPU and Monte Carlo (MC) data for the 78-site structure shown in Fig.~\ref{fig:graphene_model}\textbf{c}.
Panel \textbf{a} shows the average nitrogen concentration as a function of $\Delta\mu$  (bottom $x$-axis) and $\Delta_g$ (top $x$-axis).
Black dots and lines indicate QPU data obtained at $R_{\mathrm{NN}} = 4\,\mu$m, while coloured curves correspond to MC results at $T = 41\,\mu$K for increasing sample sizes.
Panel \textbf{b} compares the sampling frequency of the annealer (orange) and of the MC (blue) as a function of energy for $\Delta_g = 0$~eV.
Note that the $y$-axes have different orders of magnitude.
Panel \textbf{c} shows the fraction of sampled configurations as a function of nitrogen concentration for the QPU (orange bars) and the MC (blue bars) at $\Delta_g = 0$~eV. Both datasets are restricted to configurations with $E = 0$ and normalised separately.}}
    \label{fig:monte_carlo}
\end{figure}

Since the benchmark in the exhaustive search section shows good agreement between the annealing results and the exhaustive sampling for the smaller structure, we now turn to the larger system depicted in Fig.~\ref{fig:graphene_model}\textbf{c}. This structure contains 78 sites and the number of possible configurations within the composition range of interest exceeds $10^{21}$, rendering exhaustive enumeration computationally intractable. {\color{black}We therefore employ random Monte Carlo (MC) sampling as a classical reference. For each of the $10^8$ samples, a nitrogen count $N$ is drawn uniformly from 0 to 78 and $N$ sites are then selected at random, so that all compositions are explored with equal probability.} We generate a total of $10^8$ random configurations. For each configuration, the energy is computed using Eq.~\ref{eq:energies_es_2}, and the average concentration is evaluated according to Eq.~\ref{eq:average_conc_hardware}, with probabilities defined in Eq.~\ref{eq:probabilities_classical} using $T' = \alpha_v T$ where $T=41\mu$K as determined in the exhaustive search section. 
In Fig.~\ref{fig:monte_carlo}\textbf{a}, we compare the average nitrogen concentration from both the MC and QPU as a function of $\Delta_g$ (top $x$-axis) and $\Delta\mu$ (bottom $x$-axis).

We first consider the regime $\Delta_g > 0$. In this regime, a competition arises between the negative on-site term (see Eq.~\ref{eq:state_evolution_2}), which favours nitrogen incorporation, and the energy penalty associated with neighbouring nitrogen atoms. We find that as the MC sample size increases, configurations with higher nitrogen concentrations become more frequent. At the largest $\Delta_g$ value we benchmark, increasing the MC sample size drives the average concentration toward the QPU result. {\color{black}This trend highlights the difference between random sampling and the distribution generated by the QPU, which preferentially samples low-energy configurations that might be characterised by higher concentrations of nitrogen.}
%
We further find that the MC estimates are systematically lower than the QPU results, with the discrepancy shrinking as $\Delta_g$ decreases.
An interesting point arises at $\Delta_g = 0$. In this case, the presence of a nitrogen atom provides no energetic advantage, while configurations containing interacting nitrogen pairs incur an energy penalty. As a result, the lowest achievable energy is zero, which occurs either when no nitrogen atoms are present or when they are placed sufficiently far apart to be considered non-interacting. 
{\color{black}This accentuates the trend already observed for $\Delta_g \gtrsim 0$, and regardless of the sample size, no increase in the calculated average composition is observed with increasing MC sample size.}
Structures containing a larger number of non-interacting nitrogen atoms are rare at any given composition and are therefore difficult to capture using random MC sampling. In contrast, the quantum annealer explores only the low-lying energy landscape. {\color{black}Within this restricted region, the factorial growth of multiplicity makes high–nitrogen configurations more likely to appear, in accordance with probabilities expected from a Boltzmann-like distribution obtained through exhaustive search.}

{\color{black}To verify this interpretation, we examine the distribution of sampled configurations as a function of nitrogen concentration at $\Delta_g = 0$, comparing the QPU and MC results in Fig.~\ref{fig:monte_carlo}\textbf{c}. Both datasets are restricted to configurations with $E = 0$ and normalised separately, so that the comparison reflects how each method populates the zero-energy landscape across different compositions.

In the MC data, the fraction is approximately uniform for low nitrogen counts ($N = 0$--$2$, each $\sim$14\%) and decreases steadily for higher concentrations. This profile reflects the sampling scheme: because each composition is drawn with equal probability, the fraction at a given $N$ is determined by the probability that a random placement of $N$ atoms avoids all nearest-neighbour contacts. For $N = 0$ and $N = 1$ this probability is unity, while it declines with increasing $N$.

In contrast, the QPU output is concentrated at intermediate nitrogen counts, peaking at $N = 5$ (22\%) with substantial weight on $N = 3$--$7$. These configurations correspond to non-interacting arrangements that are combinatorially rare at each composition but are efficiently identified by the annealer through its preferential exploration of the low-energy configurations.

While classical importance-sampling techniques are known to efficiently target low-energy regions and reproduce Boltzmann statistics, the purpose of the present comparison is not to benchmark against such methods, but to compare random sampling with the distributions generated by the QPU.}

This comparison is reported in Fig.~\ref{fig:monte_carlo}\textbf{b}, which shows histograms of the energies of the configurations sampled by the MC method (blue) and by the QPU (orange). {\color{black}These correspond to sampling frequencies rather than Boltzmann-weighted probability distributions.} For clarity, both distributions are shown with truncated $y$-axes, as the dominant peak at $E = 0$ would otherwise obscure the remaining features. In addition, the vertical scale of the QPU distribution is an order of magnitude larger than that of the classical one; the corresponding peak heights are explicitly reported in the figure.

Only a small fraction of the MC-generated configurations, approximately 7.3\% of the total, have zero energy, while the remaining structures are broadly distributed across the accessible energy range. In contrast, the QPU output is sharply localised at low energies: about 82\% of the configurations returned by the annealer lie exactly at $E = 0$, with a secondary, much smaller peak around $E = 1 \times 10^{-6}$~eV. 

{\color{black}This pronounced accumulation at low energies shows that the QPU preferentially samples energetically favourable configurations, consistent with Boltzmann-like statistics. If we were to perform an exhaustive enumeration of all possible configurations, we would expect to recover distributions consistent with the quantum hardware data, as observed for the 28-atom system.}

{\color{black}The distributions observed in Figs.~\ref{fig:exhaustive_search} and~\ref{fig:monte_carlo} are consistent with a Boltzmann-like form that can be characterised by an effective temperature. While we do not explicitly extract a unique effective temperature across all parameter settings, this description appears to provide a useful and relatively stable characterisation of the sampled ensemble within the regimes explored here.
}

\subsection*{Tuning the Effective Temperature through Atomic Spacing}
\begin{figure}
    \centering    \includegraphics[width=0.75\linewidth]{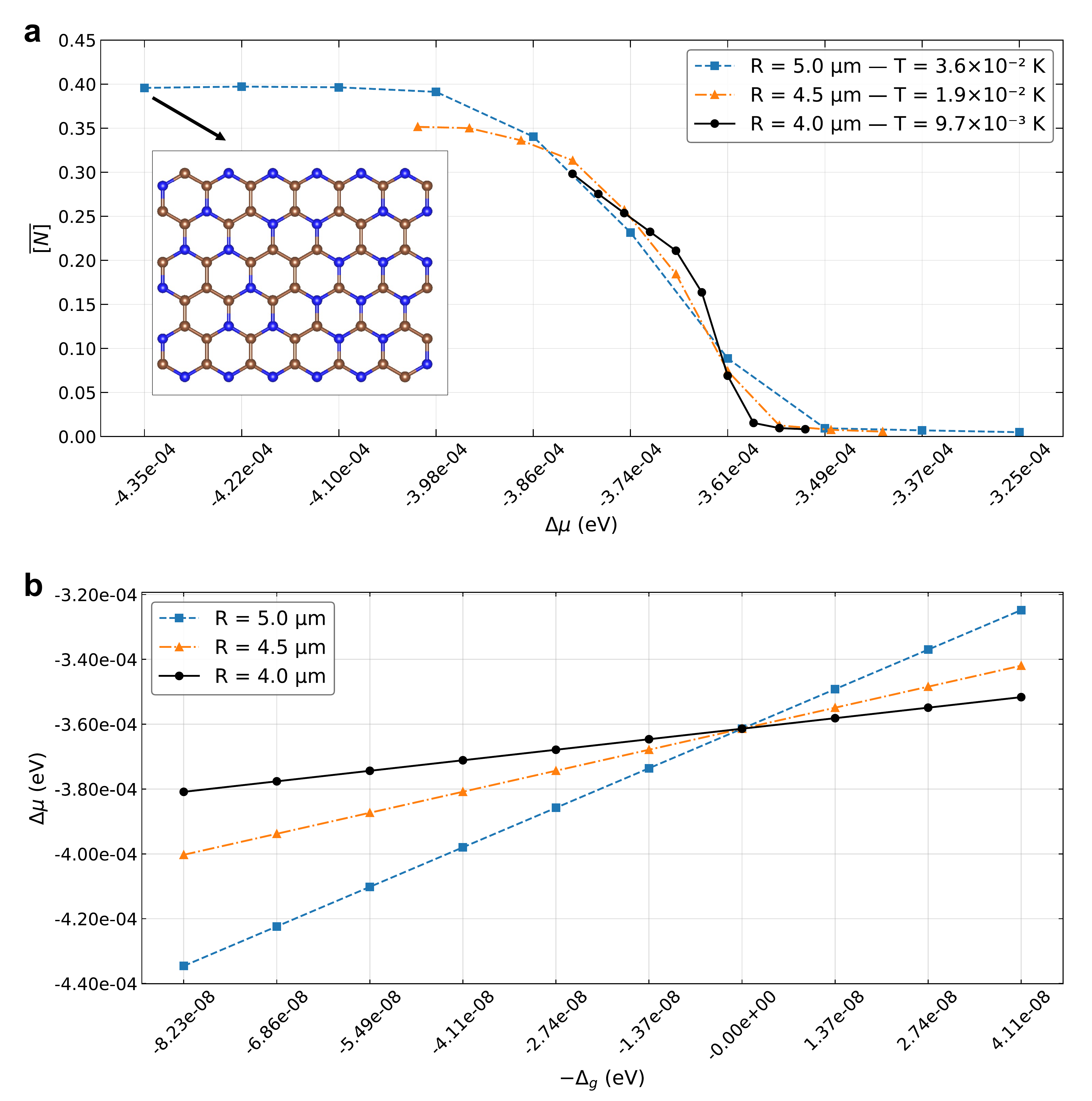}
    \caption{Panel \textbf{a} shows the average nitrogen concentration in the 78-site graphene structure depicted in Fig.~\ref{fig:graphene_model}\textbf{c} as a function of the chemical potential $\Delta\mu$ at different interatomic distances in the hardware. The inset in this panel shows an example of structure returned by the annealer in the plateau region of the $R_{\mathrm{NN}}=5\mu$m (blue) line. Panel \textbf{b} displays how the chemical potential varies as a function of the global detuning $\Delta_g$ for the three interatomic separations obtained according to Eq.~\ref{eq:def_Delta_mu^alpha_v}}
    \label{fig:graphene_diff_T}
\end{figure}

In the exhaustive search and Monte Carlo sections, we focussed on how the average nitrogen concentration depends on the chemical potential. To investigate how it depends on temperature, we use the fact that the effective temperature in the mapped doped-graphene model can be tuned by varying the nearest-neighbour spacing in the Rydberg array. As shown in Eq.~\ref{eq:def_rescaled_temperature}, the simulated temperature of the doped-graphene model is $T' = T\,\alpha_v$ where $\alpha_v=\left(R_{\mathrm{NN}}/{R_{\mathrm{NN}}^{\mathrm{DFT}}}\right)^6$. In the previous experiments, we fixed $R_{\mathrm{NN}}=R_{\mathrm{NN}}^{\mathrm{min}}$, the minimum separation supported on the QPU. Here, we increase the distance between nearest neighbours to simulate different effective temperatures.
In Fig.~\ref{fig:graphene_diff_T}\textbf{a}, we report the nitrogen concentration distributions obtained for nearest-neighbour separations of $R_{\mathrm{NN}} = 4\,\mu$m, $4.5\,\mu$m, and $5\,\mu$m, corresponding to effective temperatures of $9.7\times10^{-3}$~K, $1.9\times10^{-2}$~K, and $3.6\times10^{-2}$~K, respectively. These results illustrate how increasing the interatomic spacing on the hardware provides direct control over the effective sampling temperature.

As discussed above and formalised in Eq.~\ref{eq:def_Delta_mu^alpha_v}, changing the interatomic distance also modifies the mapping between the global detuning $\Delta_g$ and the corresponding chemical potential $\Delta\mu$. As a result, the three curves in Fig.~\ref{fig:graphene_diff_T}\textbf{a} do not span the same intervals of $\Delta\mu$. This effect is shown explicitly in Fig.~\ref{fig:graphene_diff_T}\textbf{b}, where we plot $\Delta\mu$ as a function of $\Delta_g$ for the three interatomic separations. Increasing $R_{\mathrm{NN}}$ broadens the range of chemical potentials accessed for the same global detuning window.

For $R_{\mathrm{NN}} = 5\,\mu$m, the nitrogen concentration reaches a plateau at large negative values of $\Delta\mu$. In this regime, the configurations returned by the annealer correspond to structures that maximise nitrogen incorporation while avoiding nearest-neighbour nitrogen pairs. An example of such a configuration is shown in the inset of Fig.~\ref{fig:graphene_diff_T}\textbf{a}. At this separation, the nearest-neighbour interaction energy between two simultaneously excited atoms is $2.28\times10^{-7}$~eV, which exceeds the maximum detuning available on the current QuEra device. Consequently, within the experimentally accessible range of $\Delta_g$, configurations containing adjacent nitrogen atoms are suppressed for all interatomic distances considered here. We therefore expect that access to larger detuning values would enable sampling of configurations with higher nitrogen concentrations.

This result is particularly significant: in addition to demonstrating that the laser detuning in the quantum annealer can be directly mapped onto the chemical potential in the material, we show that Boltzmann-like distributions at different effective temperatures can be realised simply by tuning the interatomic distance in the hardware. This establishes a direct and experimentally accessible means of multivariate thermodynamic control within the quantum annealer.

\section*{Discussion}
\label{sec:conclusions}

In this work, we presented a workflow for mapping the DFT-derived energetics of nitrogen-doped graphene onto a Rydberg Hamiltonian, enabling quantum annealing on the QuEra Aquila hardware. The mapping expresses the formation energies in terms of an on-site term (implemented through the global detuning) and pairwise Van der Waals interactions encoded through the programmable interatomic separations of the neutral-atom array.

A central component of this approach is the introduction of a rescaling factor, $\alpha_v$, which accounts for hardware constraints while preserving the thermodynamic structure of the problem. We showed analytically that this rescaling leads to an effective annealing temperature $T' = \alpha_v T$, allowing the sampled distributions to be interpreted in thermodynamic terms even when the exact DFT interaction strengths cannot be realised on the hardware.

The approach was first validated on a 28-site graphene nanoflake through exhaustive enumeration, and then applied to a 78-site system using {\color{black}random} Monte Carlo sampling. In both cases, the quantum annealer reproduces the expected thermodynamic trends despite the large configuration space. We also showed that the effective temperature can be tuned experimentally through the interatomic separation, providing a direct handle on the sampled distributions.

{\color{black}While classical importance-sampling methods remain highly effective in studying the thermodynamic properties of materials, in this work, we develop a mapping from materials models to a Rydberg Hamiltonian and show, as a proof of concept, that the resulting quantum hardware output can be interpreted in thermodynamic terms. This establishes a basis for the development of quantum-ready models and algorithms for materials problems.

Within this framework, the main limitations arise from the form of the Hamiltonian available on the hardware. In particular, the restriction to two-body interactions and the fixed $1/r^6$ dependence make it difficult to capture systems with strong localised interactions or many-body effects, such as the phosphorous-nitrogen co-doping case discussed in the Methods section. Extending this approach to such systems will require more flexible interaction models or hardware capable of implementing higher-order couplings.}

\section*{Methods}
\label{sec:methods}

\subsection*{Materials and DFT Calculation Details}
\label{sec:dft_details}
Density functional theory (DFT) calculations for nitrogen-doped graphene were performed using the \textsc{CRYSTAL23} code~\cite{crystal23}, which constructs crystalline orbitals from localised Gaussian basis functions. The exchange–correlation functional was treated within the PBE approximation~\cite{Perdew1996}, employing the revised pob-TZVP basis set~\cite{Vilela2019}, which includes polarisation functions for all elements. 

The truncation of the Coulomb and exchange lattice series in \textsc{CRYSTAL23} is governed by five thresholds, set to 8 (T1–T4) and 16 (T5). The self-consistent field (SCF) convergence criteria were $10^{-8}$~Hartree for total energy and $10^{-6}$~Hartree for structural relaxation. Reciprocal-space integration employed a Pack–Monkhorst grid centred at the $\Gamma$ point, with shrinking factors of 6 and 12 for geometry optimisation and energy evaluation, respectively. 

The graphene supercell used in these calculations contains 78 atomic sites (Fig.~\ref{fig:graphene_model}\textbf{a}). This cell was selected because it naturally accommodates the nanoflake structure shown in Fig.~\ref{fig:graphene_model}\textbf{c}. Nitrogen dopants were introduced in concentrations ranging from one to ten atoms per cell, corresponding to 1.28\%–12.8\% nitrogen.

For the DFT dataset, we begin by generating configurations at random while enforcing that no two structures are related by the symmetries of the pristine graphene supercell. From this pool of symmetry-inequivalent configurations (SICs), we select a subset of size $N_{\text{train}}^{\text{SIC}}$ for explicit DFT evaluation. Each chosen SIC is relaxed at constant pressure, and its energy is then assigned to all symmetry-equivalent configurations (SECs) obtained by applying the space group operations of the underlying graphene lattice. {\color{black}The SECs do not correspond to additional DFT calculations; they are included in the regression so that all symmetry-equivalent sites receive the same fitted parameters in the Rydberg Hamiltonian.} In this way, a set of $N_{\text{train}}^{\text{SEC}}$ labelled structures is generated and used to fit the Rydberg Hamiltonian parameters. The same symmetry-based expansion is used for the test set. {\color{black}The distribution of SICs and SECs across nitrogen concentrations is reported in Supplementary Table~S1.}

Formation energies were referenced to molecular nitrogen and pristine graphene, ensuring that the energy of pure graphene is set to zero. Because \textsc{CRYSTAL23} employs localised basis sets, the graphene calculations are periodic in two dimensions, while those for the nitrogen molecule are treated as zero-dimensional, without the need for vacuum separation.

The energies discussed above were calculated for a two-dimensional periodic layer of graphene, as illustrated in Fig.~\ref{fig:graphene_model}\textbf{a}. This approach simplifies the mapping of the DFT-derived energies onto the Rydberg Hamiltonian. However, in the annealing experiments we employ finite, non-periodic nanoflakes, shown in Fig.~\ref{fig:graphene_model}\textbf{b} and ~\ref{fig:graphene_model}\textbf{c}. If the energies had been calculated directly for these non-periodic structures, slight variations would have appeared in the on-site energies and two-body interactions of atoms located at the edges, as these atoms have only two nearest neighbours instead of three, unlike those in the centre of the cell.
In principle, the resulting variations in on-site energies could be compensated by using site-dependent local detuning. However, the corresponding differences in $V_{i,j}$ would lead to modified atomic positions that cannot be realised on the hardware, given the geometric constraints discussed in the main text. For this reason, we extract a portion of the periodic graphene layer and treat it as representative of the bulk material.

This approximation has a clear physical consequence: nitrogen atoms tend to occupy boundary sites more frequently than they would in a fully periodic model, where atoms at opposite edges are nearest neighbours. The energetic implication is that, in the non-periodic case, the energy associated with two distant nitrogen atoms is $2V^{\mathrm{DFT}}$, whereas in the periodic system it is $2V^{\mathrm{DFT}} + V_{\mathrm{NN}}^{\mathrm{DFT}}$. As a result, configurations with nitrogen atoms at the boundaries are slightly overrepresented during thermodynamic sampling compared with those in the centre of the flake. Nevertheless, the number of such configurations is small relative to the total, and their contribution becomes negligible as the system size increases. 

Unlike D-Wave annealers~\cite{Camino2023,Camino2025}, the QuEra architecture does not natively support periodic boundary conditions. In principle, opposite sides of the cell could be connected through additional “gadgets,” as proposed in Ref.~\cite{Nguyen2023}, but this approach would introduce substantial hardware overhead. For this reason, we did not pursue this route in the present work. However, periodic boundary conditions could be achieved, in principle, with a reconfigurable atom array.

\subsection*{Quantum Annealing Details}
\label{sec:qa_details}
Quantum annealing was performed on the QuEra Aquila~\cite{Wurtz2023} device. These employes Rb-87 atoms are held and cooled to microkelvin temperatures by laser beams inside a vacuum cell. Reconfigurable optical tweezers enable the arbitrary arrangements of up to 256 atoms. The annealing is carried out by lasers that excite the Rb-87 atoms to their Rydberg state where they interact with each other.

\begin{figure}
    \centering    \includegraphics[width=0.75\linewidth]{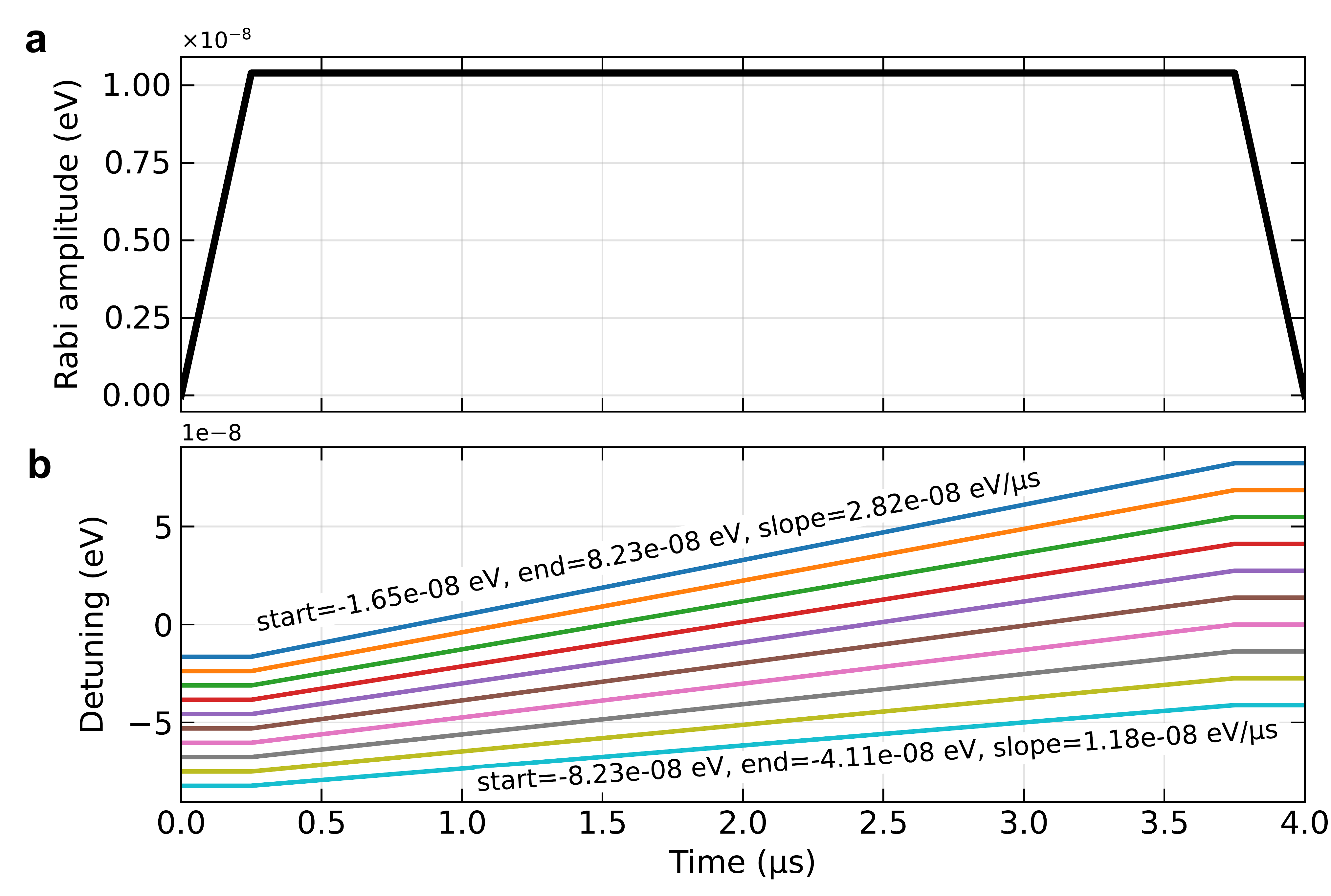}
    \caption{Laser energies used in the experiments reported in this work. Panel \textbf{a} depicts the Rabi amplitude values throughout the annealing. Panel \textbf{b} shows the detuning values variations during the annealing for the ten different values of final $\Delta_g$ used in the Results sections.}
    \label{fig:global_drive}
\end{figure}

In our experiments, two types of laser pulses are applied during the annealing process, as illustrated in Fig.~\ref{fig:global_drive}. Panel \textbf{a} of the figure shows the Rabi drive, which is kept identical across all experiments, while the \textbf{b} panel depicts the global detuning, whose slope varies depending on its final value. Ideally, the detuning profile should remain consistent between experiments; to achieve this, we implemented a piecewise-linear detuning scheme. 
Each detuning pulse was divided into three regions: an initial and a final hold, each spanning 6.25\% of the total sequence duration, and a central 87.5\% linear ramp connecting the starting and ending detuning values. The total pulse duration was fixed at $t = 4~\mu\text{s}$. The final detuning values were chosen to decrease linearly from $-8.23\times10^{-8}$~eV to $4.11\times10^{8}$~eV. The corresponding initial detuning values were clipped at the lower hardware limit of $8.23\times10^{-8}$~eV. 
This construction yields ramp rates between $2.82 \times10^{-8}$~eV/$\mu$s and $1.18 \times10^{-8}$~eV/$\mu$s while maintaining identical pulse shapes and timing across all experiments. This method ensures moderate slope variation within the detuning window, preventing abrupt transitions while covering a large fraction of the available dynamic range.
For each data point, the annealing protocol was executed 1000 times in order to obtain statistically meaningful thermodynamic averages. Prior to annealing, the rubidium atoms are arranged in the target geometry; however, due to experimental imperfections, not all sites are occupied in every run. 
{\color{black}To ensure consistency, we retain only those annealing outcomes corresponding to fully occupied initial configurations. Across all experiments reported in this work, the fraction of runs with complete initial occupancy averaged 60.5\%, with a standard deviation of 12.1\%. The observed occupancy fraction ranged from a minimum of 40.6\% to a maximum of 81.4\%, depending on the experiment. This filtering reduces the number of samples but does not introduce a systematic bias, and therefore does not affect the comparison with the reference distributions.}

\subsection*{Phosphorous–Nitrogen Co-doping: Challenges for the Current Mapping}
\label{P-N_doping}
In this section, we briefly examine an extension of the nitrogen-doped graphene model in which a single phosphorous atom is introduced at a fixed lattice site. The system is described using the 78-site graphene cell shown in Fig.~\ref{fig:graphene_model}\textbf{c}, with the phosphorous atom placed at the centre of the structure and held fixed throughout the analysis. This case is included to illustrate the challenges that arise when additional chemical complexity is introduced, and to clarify the limitations of the current mapping strategy when applied to systems with strong, localised interactions.

\begin{figure}
    \centering
    \includegraphics[width=0.5\linewidth]{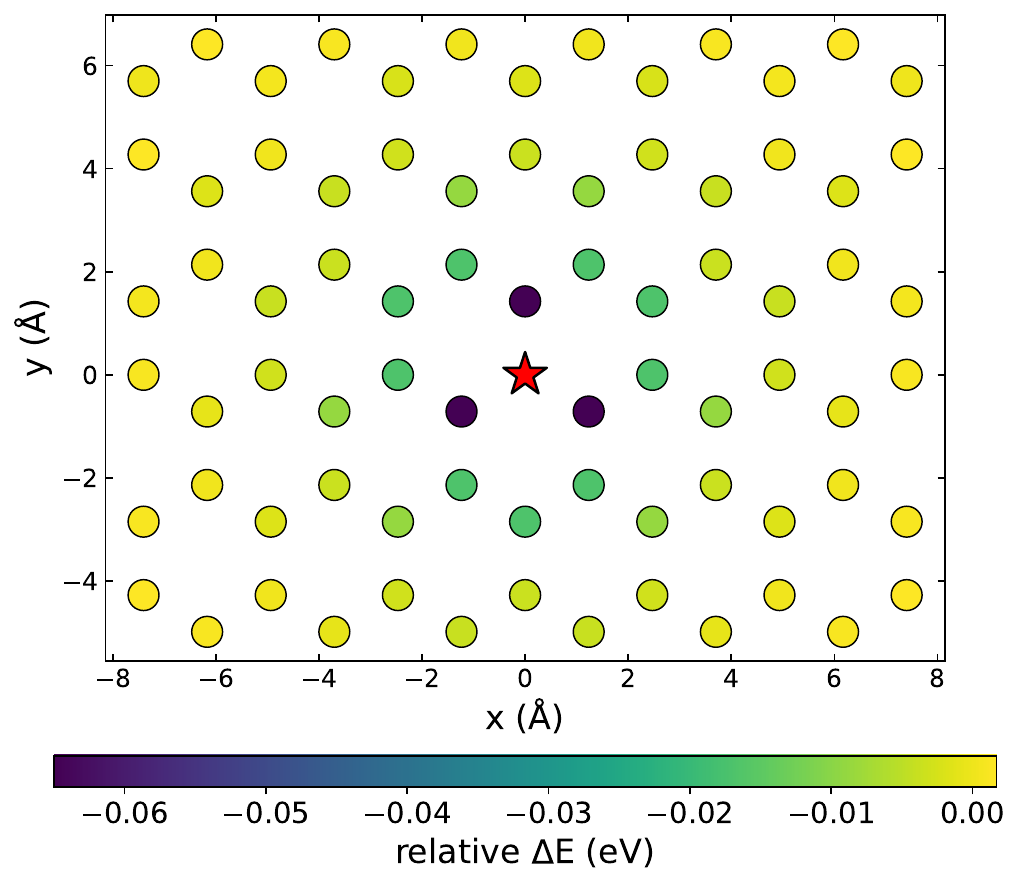}
    \caption{Graphene structure containing 78 sites used to construct the energy model for phosphorous-doped graphene. The red star marks the position of the phosphorous atom. Circles denote the carbon sites, and the heat-map colour scale indicates the energy associated with placing a nitrogen atom at each site relative to placing it at the maximum separation, where the nitrogen–phosphorous interaction is negligible.}
    \label{fig:graphene_B}
\end{figure}

For this problem we make one additional modelling assumption: we retain the two-body nitrogen–nitrogen interaction energy obtained in Eq.~\ref{eq:linear_regression_ryd}, and allow only the on-site energy of nitrogen to vary in order to capture its preference to sit closer or further from the phosphorous atom. To obtain this distance-dependent on-site term, a single nitrogen atom is placed, one site at a time, at each symmetry-inequivalent distance from the phosphorous atom. The resulting eighteen structures are relaxed and their energies computed using DFT. The on-site potential at each distance is then defined relative to a reference configuration in which the nitrogen atom is placed at the maximum separation allowed by the cell (9.34~\AA), where the interaction with phosphorous is considered to be negligible. The resulting distance-dependent on-site energies are shown in Fig.~\ref{fig:graphene_B}. As expected, nitrogen strongly favours occupying a site adjacent to phosphorous (energy gain of $-1.11$~eV). 
This behaviour reflects the distinct atomic radius and electronegativity of the two dopants. Phosphorus has a significantly larger covalent radius than carbon and nitrogen, leading to local lattice distortion when incorporated into the graphene network. Nitrogen, by contrast, is smaller and more electronegative, and preferentially stabilises regions of high local strain and charge redistribution. When placed adjacent to phosphorus, nitrogen partially compensates the local structural and electronic perturbation induced by the larger dopant, resulting in a net stabilisation of the local bonding environment and a lower formation energy.

This chemical preference introduces the first of two challenges when mapping this system to the quantum annealer. The energy scale associated with the nitrogen–phosphorous interaction is approximately four orders of magnitude larger than the on-site term $V^{\mathrm{DFT}}$ extracted from the nitrogen–nitrogen problem. Since the local detuning on the hardware has the same energy scale as the global detuning, even after applying the rescaling procedure introduced above, the local term would dominate and suppress any global-detuning driven sampling of compositions.
The second challenge is that the effective interactions are no longer strictly two-body. The strong stabilisation of a nitrogen atom placed next to phosphorus is specific to the first dopant and arises from a combination of local lattice relaxation due to the larger atomic radius of phosphorus and the electronegativity contrast between the two species. Once one nitrogen atom occupies the site adjacent to phosphorous, the energy gain associated with placing a second nitrogen nearby decreases to $-6.35\times10^{-1}$~eV (about 57\% smaller), and to $-2.86\times10^{-1}$~eV (26\% smaller) when a third is added. Thus, the energetics depend on the number of nitrogen atoms already present in the local environment. These higher-order many-body effects cannot be captured by a purely two-body Hamiltonian.
In principle, this behaviour could be approximated by averaging the distance-dependent interaction energies, but such an approach would obscure the many-body nature of the problem. A more rigorous treatment would require a Hamiltonian capable of representing higher-order interactions, such as a Higher-Order Unconstrained Binary Optimisation (HUBO) model. While current Rydberg-atom hardware natively supports only two-body interactions, future architectures incorporating multi-body couplings may enable an exact mapping of this class of chemically complex systems.

\section*{Supplementary Information}

\section*{Data Availability}
\label{sec:data_availability}
All data supporting the findings of this study are available at https://github.com/cmc-ucl/rydberg\_atoms.

\section*{Acknowledgements}
\label{sec:funding_declaration}
This work was supported by the UKRI EPSRC Grant No. EP/X035859 and EPSRC Grant No. EP/W026775. Via our membership of the UK's HEC Materials Chemistry Consortium, which is funded by EPSRC (EP/W026775), this work used the ARCHER2 UK National Supercomputing Service (http://www.archer2.ac.uk). 

\section*{Competing Interests}
The authors declare no competing interests.

\newpage
\bibliography{sn-bibliography}

@article{Allan2018,
	Author = {Allan, N. L. and Thomas, L. and Hart, J. N. and Freeman, C. L. and Mohn, C. E.},
	Da = {2019/02/01},
	Doi = {10.1007/s00269-018-0997-3},
	Id = {Allan2019},
	Isbn = {1432-2021},
	Journal = {Physics and Chemistry of Minerals},
	Number = {2},
	Pages = {193--202},
	Title = {Calcite--magnesite solid solutions: using genetic algorithms to understand non-ideality},
	Volume = {46},
	Year = {2019}
    }

@article{Barredo2018,
  author  = {Barredo, Daniel and Lienhard, Vincent and de L{\'e}s{\'e}leuc, Sylvain and Lahaye, Thierry and Browaeys, Antoine},
  title   = {Synthetic three-dimensional atomic structures assembled atom by atom},
  journal = {Nature},
  volume  = {561},
  number  = {7721},
  pages   = {79--82},
  year    = {2018},
  doi     = {10.1038/s41586-018-0450-2}
}

@article{Binninger2025,
  title = {Simulating charging characteristics of lithium iron phosphate by electro-ionic optimization on a quantum annealer},
  author = {Binninger, Tobias and Ting, Yin-Ying and K\"oster, Konstantin and Bruch, Nils and Kaghazchi, Payam and Kowalski, Piotr M. and Eikerling, Michael H.},
  journal = {Phys. Rev. B},
  volume = {112},
  issue = {17},
  pages = {174118},
  numpages = {15},
  year = {2025},
  month = {Nov},
  publisher = {American Physical Society},
  doi = {10.1103/cpgy-fpvb}
}

@article{Bluvstein2022,
  author       = {Bluvstein, Dolev and Levine, Harry and Semeghini, Giulia and Wang, Tout T. and Ebadi, Sepehr and Kalinowski, Marcin and Keesling, Alexander and Maskara, Nishad and Pichler, Hannes and Greiner, Markus and Vuleti{\'c}, Vladan and Lukin, Mikhail D.},
  title        = {A quantum processor based on coherent transport of entangled atom arrays},
  journal      = {Nature},
  year         = {2022},
  volume       = {604},
  number       = {7906},
  pages        = {451--456},
  doi          = {10.1038/s41586-022-04592-6}
}

@article{Browaeys2020,
  author  = {Browaeys, Antoine and Lahaye, Thierry},
  title   = {Many-Body Physics with Individually Controlled {Rydberg} Atoms},
  journal = {Nature Physics},
  year    = {2020},
  volume  = {16},
  number  = {2},
  pages   = {132--142},
  doi     = {10.1038/s41567-019-0733-z},
  issn    = {1745-2481}
}

@article{Camino2023,
    author = {Camino, B. and Buckeridge, J. and Warburton, P. A. and Kendon, V. and Woodley, S. M.},
    title = {Quantum computing and materials science: A practical guide to applying quantum annealing to the configurational analysis of materials},
    journal = {Journal of Applied Physics},
    volume = {133},
    number = {22},
    pages = {221102},
    year = {2023},
    month = {06},
    issn = {0021-8979},
    doi = {10.1063/5.0151346}
}

@article{Camino2025,
author = {Bruno Camino  and John Buckeridge  and Nicholas Chancellor  and C. Richard A. Catlow  and Anna Maria Ferrari  and Paul A. Warburton  and Alexey A. Sokol  and Scott M. Woodley },
title = {Exploring the thermodynamics of disordered materials with quantum computing},
journal = {Science Advances},
volume = {11},
number = {23},
pages = {eadt7156},
year = {2025},
doi = {10.1126/sciadv.adt7156}}

@article{Scholl2021,
  author  = {Scholl, P. and Schuler, M. and Williams, H. J. and Eberharter, A. A. and Barredo, D. and Schymik, K.-N. and Lienhard, V. and Henry, L.-P. and Lang, T. C. and Lahaye, T. and L{\"a}uchli, A. M. and Browaeys, A.},
  title   = {Quantum simulation of 2D antiferromagnets with hundreds of {Rydberg} atoms},
  journal = {Nature},
  volume  = {595},
  pages   = {233--238},
  year    = {2021},
  doi     = {10.1038/s41586-021-03585-1}
}

@article{Choubisa2023,
title = {Accelerated chemical space search using a quantum-inspired cluster expansion approach},
journal = {Matter},
volume = {6},
number = {2},
pages = {605-625},
year = {2023},
issn = {2590-2385},
doi = {https://doi.org/10.1016/j.matt.2022.11.031},
author = {Hitarth Choubisa and Jehad Abed and Douglas Mendoza and Hidetoshi Matsumura and Masahiko Sugimura and Zhenpeng Yao and Ziyun Wang and Brandon R. Sutherland and Alán Aspuru-Guzik and Edward H. Sargent}
}

@article{Zhang1991,
  title = {Chemical potential dependence of defect formation energies in {GaAs}: Application to {Ga} self-diffusion},
  author = {Zhang, S. B. and Northrup, John E.},
  journal = {Phys. Rev. Lett.},
  volume = {67},
  issue = {17},
  pages = {2339--2342},
  numpages = {0},
  year = {1991},
  month = {Oct},
  publisher = {American Physical Society},
  doi = {10.1103/PhysRevLett.67.2339},
}

@article{Jeong2025,
author = {Jeong, Seokho and Park, Juyoung and Ahn, Jaewook},
title = {Quantum-Enhanced Simulated Annealing Using {Rydberg} Atoms},
journal = {Advanced Quantum Technologies},
volume = {8},
number = {12},
pages = {e2500070},
doi = {https://doi.org/10.1002/qute.202500070},
year = {2025}
}

@article{Wurtz2024,
      title={Solving non-native combinatorial optimization problems using hybrid quantum-classical algorithms}, 
      author={Jonathan Wurtz and Stefan Sack and Sheng-Tao Wang},
      year={2024},
      archivePrefix={arXiv},
      primaryClass={quant-ph},
      doi={10.48550/arXiv.2403.03153}
}

@article{Lotshaw2023,
  title = {Approximate {Boltzmann} distributions in quantum approximate optimization},
  author = {Lotshaw, Phillip C. and Siopsis, George and Ostrowski, James and Herrman, Rebekah and Alam, Rizwanul and Powers, Sarah and Humble, Travis S.},
  journal = {Phys. Rev. A},
  volume = {108},
  issue = {4},
  pages = {042411},
  numpages = {14},
  year = {2023},
  month = {Oct},
  publisher = {American Physical Society},
  doi = {10.1103/PhysRevA.108.042411}
}

@article{Ebadi2022,
author = {S. Ebadi  and A. Keesling  and M. Cain  and T. T. Wang  and H. Levine  and D. Bluvstein  and G. Semeghini  and A. Omran  and J.-G. Liu  and R. Samajdar  and X.-Z. Luo  and B. Nash  and X. Gao  and B. Barak  and E. Farhi  and S. Sachdev  and N. Gemelke  and L. Zhou  and S. Choi  and H. Pichler  and S.-T. Wang  and M. Greiner  and V. Vuletić  and M. D. Lukin },
title = {Quantum optimization of maximum independent set using {Rydberg} atom arrays},
journal = {Science},
volume = {376},
number = {6598},
pages = {1209-1215},
year = {2022},
doi = {10.1126/science.abo6587}}

@article{crystal23,
author = {Erba, Alessandro and Desmarais, Jacques K. and Casassa, Silvia and Civalleri, Bartolomeo and Donà, Lorenzo and Bush, Ian J. and Searle, Barry and Maschio, Lorenzo and Edith-Daga, Loredana and Cossard, Alessandro and Ribaldone, Chiara and Ascrizzi, Eleonora and Marana, Naiara L. and Flament, Jean-Pierre and Kirtman, Bernard},
title = {CRYSTAL23: A Program for Computational Solid State Physics and Chemistry},
journal = {Journal of Chemical Theory and Computation},
volume = {0},
number = {0},
pages = {null},
year = {0},
doi = {10.1021/acs.jctc.2c00958},
note ={PMID: 36502394}
}

@misc{Dwave, 
title={D-Wave systems}, 
url={https://www.dwavesys.com/}, 
note = {{https://www.dwavesys.com/}} }

@article{Henriet2020,
  doi = {10.22331/q-2020-09-21-327},
  title = {Quantum computing with neutral atoms},
  author = {Henriet, Lo{\"{i}}c and Beguin, Lucas and Signoles, Adrien and Lahaye, Thierry and Browaeys, Antoine and Reymond, Georges-Olivier and Jurczak, Christophe},
  journal = {{Quantum}},
  issn = {2521-327X},
  publisher = {{Verein zur F{\"{o}}rderung des Open Access Publizierens in den Quantenwissenschaften}},
  volume = {4},
  pages = {327},
  month = sep,
  year = {2020}
}

@inbook{Kochenberger2025,
  author    = {Kochenberger, Gary A. and Glover, Fred and Wang, Haibo},
  editor    = {Pardalos, Panos M. and Du, Ding-Zhu and Thai, My T.},
  title     = {{QUBO}: Quadratic Unconstrained Binary Optimization Problem},
  booktitle = {Handbook of Combinatorial Optimization},
  year      = {2025},
  publisher = {Springer New York},
  address   = {New York, NY},
  pages     = {1--26},
  isbn      = {978-1-4614-6624-6},
  doi       = {10.1007/978-1-4614-6624-6\_15-1}
}

@article{Lucas2014,
  author    = {Lucas, Andrew},
  title     = {Ising Formulations of Many {NP} Problems},
  journal   = {Frontiers in Physics},
  year      = {2014},
  volume    = {2},
  pages     = {5},
  doi       = {10.3389/fphy.2014.00005},
  issn      = {2296-424X}
}

@article{Gusev2023,
  title={Optimality guarantees for crystal structure prediction},
  author={Gusev, Vladimir V and Adamson, Duncan and Deligkas, Argyrios and Antypov, Dmytro and Collins, Christopher M and Krysta, Piotr and Potapov, Igor and Darling, George R and Dyer, Matthew S and Spirakis, Paul and others},
  journal={Nature},
  volume={619},
  number={7968},
  pages={68--72},
  year={2023},
  publisher={Nature Publishing Group UK London}
}

@article{Mohn2015,
doi = {10.1088/0953-8984/27/42/425201},
year = {2015},
month = {oct},
publisher = {IOP Publishing},
volume = {27},
number = {42},
pages = {425201},
author = {Mohn, Chris E and Kob, Walter},
title = {Predicting complex mineral structures using genetic algorithms},
journal = {Journal of Physics: Condensed Matter}
}

@article{Mohn2018,
author = {Chris E. Mohn},
title = {Predicting cation ordering in {MgAl$_2$O$_4$} using genetic algorithms and density functional theory},
journal = {Materials and Manufacturing Processes},
volume = {33},
number = {2},
pages = {174--179},
year = {2018},
publisher = {Taylor \& Francis},
doi = {10.1080/10426914.2017.1303153}
}

@article{Nguyen2023,
  title = {Quantum Optimization with Arbitrary Connectivity Using {Rydberg} Atom Arrays},
  author = {Nguyen, Minh-Thi and Liu, Jin-Guo and Wurtz, Jonathan and Lukin, Mikhail D. and Wang, Sheng-Tao and Pichler, Hannes},
  journal = {PRX Quantum},
  volume = {4},
  issue = {1},
  pages = {010316},
  numpages = {19},
  year = {2023},
  month = {Feb},
  publisher = {American Physical Society},
  doi = {10.1103/PRXQuantum.4.010316}
}

@article{Perdew1996,
  title = {Generalized Gradient Approximation Made Simple},
  author = {Perdew, John P. and Burke, Kieron and Ernzerhof, Matthias},
  journal = {Phys. Rev. Lett.},
  volume = {77},
  issue = {18},
  pages = {3865--3868},
  numpages = {0},
  year = {1996},
  month = {Oct},
  publisher = {American Physical Society},
  doi = {10.1103/PhysRevLett.77.3865},
}

@article{Purton2007,
title = {{Monte Carlo simulation of GaN/AlN and AlN/InN mixtures}},
journal = {Materials Chemistry and Physics},
volume = {105},
number = {2},
pages = {179-184},
year = {2007},
issn = {0254-0584},
doi = {https://doi.org/10.1016/j.matchemphys.2007.04.024},
author = {John A. Purton and Mikhail Yu. Lavrentiev and Neil L. Allan}
}

@article{Vilela2019,
author = {Vilela Oliveira, Daniel and Laun, Joachim and Peintinger, Michael F. and Bredow, Thomas},
title = {BSSE-correction scheme for consistent gaussian basis sets of double- and triple-zeta valence with polarization quality for solid-state calculations},
journal = {Journal of Computational Chemistry},
volume = {40},
number = {27},
pages = {2364-2376},
doi = {https://doi.org/10.1002/jcc.26013},
year = {2019}
}

@article{Wintersperger2023,
  author  = {Wintersperger, K. and Dang, H.-L. and Capecci, G. and Krinner, S. and Scheuer, J. and Fedorov, A. and Barredo, D. and Browaeys, A. and Kuhr, S. and Monz, T. and Schaetz, T. and Blatt, R. and Bloch, I.},
  title   = {Neutral Atom Quantum Computing Hardware: Performance and End-User Perspective},
  journal = {EPJ Quantum Technology},
  year    = {2023},
  volume  = {10},
  pages   = {32},
  doi     = {10.1140/epjqt/s40507-023-00190-1}
}

@article{Wurtz2023,
      title={Aquila: QuEra's 256-qubit neutral-atom quantum computer}, 
      author={Jonathan Wurtz and Alexei Bylinskii and Boris Braverman and Jesse Amato-Grill and Sergio H. Cantu and Florian Huber and Alexander Lukin and Fangli Liu and Phillip Weinberg and John Long and Sheng-Tao Wang and Nathan Gemelke and Alexander Keesling},
      year={2023},
      primaryClass={quant-ph},
      doi={https://doi.org/10.48550/arXiv.2306.11727}, 
}

@misc{Zhao2025,
      title={An integrated photonics platform for high-speed, ultrahigh-extinction, many-channel quantum control}, 
      author={Mengdi Zhao and Manuj Singh and Anshuman Singh and Henry Thoreen and Robert J. DeAngelo and Daniel Dominguez and Andrew Leenheer and Frédéric Peyskens and Alexander Lukin and Dirk Englund and Matt Eichenfield and Nathan Gemelke and Noel H. Wan},
      year={2025},
      eprint={2508.09920},
      archivePrefix={arXiv},
      primaryClass={quant-ph},
      doi={https://doi.org/10.48550/arXiv.2508.09920}, 
}

@article{Hohenberg1964,
  author  = {Hohenberg, P. and Kohn, W.},
  title   = {Inhomogeneous Electron Gas},
  journal = {Physical Review},
  year    = {1964},
  volume  = {136},
  number  = {3B},
  pages   = {B864--B871},
  doi     = {10.1103/PhysRev.136.B864}
}

@article{Kohn1965,
  author  = {Kohn, W. and Sham, L. J.},
  title   = {Self-Consistent Equations Including Exchange and Correlation Effects},
  journal = {Physical Review},
  year    = {1965},
  volume  = {140},
  number  = {A4},
  pages   = {A1133--A1138},
  doi     = {10.1103/PhysRev.140.A1133}
}

\end{document}